\documentclass[prfluids,10pt,superscriptaddress]{revtex4-2}

\usepackage{preamble}
\graphicspath{{./Figures/}}

\usepackage{xpatch}
\xpatchcmd\bibsection{19}{6}{}{}
\xpatchcmd\bibsection{\begingroup}{\vskip -13pt\begingroup}{}{}

\newcommand{\htil}{\tilde{h}}
\newcommand{\dhtil}{\dot{\tilde{h}}}
\newcommand{\ctil}{\tilde{c}}
\newcommand{\dctil}{\dot{\tilde{c}}}

\newcommand{\tpb}{t_{\text{pb}}}
\newcommand{\xpb}{x_{\text{pb}}}
\newcommand{\Frpb}{\Fr_{\text{pb}}}
\newcommand{\hpb}{x_{\text{pb}}}
\newcommand{\Hpb}{h_{\text{pb}}}

\newcommand{\xb}{x_{\text{b}}}
\newcommand{\hb}{h_{\text{b}}}
\newcommand{\Hb}{H_{\text{b}}}

\newcommand{\HnoWind}{H^*_u}
\newcommand{\LnoWind}{L^*_u}

\newcommand{\xpeak}{x_{\text{peak}}}
\newcommand{\tpeak}{t_{\text{peak}}}
\newcommand{\cpeak}{c_{\text{peak}}}

\newcommand{\tquarter}{t_{0.25}}

\newcommand{\Lpz}{L_{\text{pz}}}
\newcommand{\cadi}{c_{\text{adi}}}
\newcommand{\Lbz}{L_{\text{bz}}}

\DeclareMathOperator{\Fr}{Fr}
\DeclareMathOperator{\Ir}{Ir}
\DeclarePairedDelimiter\roundeven{\lfloor}{\rceil}

\renewcommand*{\epsilon}{\varepsilon}

\begin{document}

\title{Wind-induced changes to shoaling surface gravity wave shape}

\author{Thomas Zdyrski}
\email{tzdyrski@ucsd.edu}
\author{Falk Feddersen}

\affiliation{Scripps Institution of Oceanography,
University of California, San Diego,
La Jolla, California 92092-0209, USA}

\date{5 March 2022}

\begin{abstract}
Unforced shoaling waves experience growth and changes to wave shape.
Similarly, wind-forced waves on a flat-bottom likewise experience
growth/decay and changes to wave shape.
However, the combined effect of shoaling and wind-forcing on wave shape,
particularly relevant in the near-shore environment, has not yet been
investigated theoretically.
Here, we consider small-amplitude, shallow-water solitary waves
propagating up a gentle, planar bathymetry forced by a weak,
Jeffreys-type wind-induced surface pressure.
We derive a variable-coefficient Korteweg--de Vries--Burgers (vKdV--B)
equation governing the wave profile's evolution and solve it
numerically using a Runge-Kutta third-order finite difference solver.
The simulations run until convective prebreaking---a Froude number
limit appropriate to the order of the vKdV--B equation.
Offshore winds weakly enhance the ratio of prebreaking height to depth
as well as prebreaking wave slope.
Onshore winds have a strong impact on narrowing the wave peak, and
wind also modulates the rear shelf formed behind the wave.
Furthermore, wind strongly affects the width of the prebreaking zone,
with larger effects for smaller beach slopes.
After converting our pressure magnitudes to physically realistic wind
speeds, we observe qualitative agreement with prior laboratory and
numerical experiments on breakpoint location.
Additionally, our numerical results have qualitatively similar temporal
wave shape to shoaling and wind-forced laboratory observations,
suggesting that the vKdV--B equation captures the essential aspects of
wind-induced effects on shoaling wave shape.
Finally, we isolate the wind's effect by comparing the wave profiles to
the unforced case.
This reveals that the numerical results are approximately a
superposition of a solitary wave, a shoaling-induced shelf, and a
wind-induced, bound, dispersive, and decaying tail.
\end{abstract}


\maketitle

\section{Introduction}
\noindent
Wind coupled to surface gravity waves leads to wave growth and decay
as well as changes to wave shape.
However, many aspects of wind-wave coupling are not yet fully
understood.
Since the sheltering theory of wind-wave coupling by
\citet{jeffreys1925formation}, a variety of mechanisms for wind-wave
interactions have been put forward, often with a focus on calculating
growth rates~\citep{miles1957generation,phillips1957generation}.
Furthermore, these theories have been tested by many studies in the
laboratory~\citep{wu1968laboratory,phillips1974wave,plant1977growth,%
buckley2019turbulent} and the
field~\citep{hasselmann1973measurements,donelan2006wave}.
Similarly, numerical studies modeled the airflow above waves using
methods such as large eddy
simulations~\citep{hara2015wave,yang2013dynamic,husain2019boundary}
or modeled the combined air and water domain using Reynolds-averaged
Navier--Stokes (RANS) solvers~\citep{zou2017wind} or direct numerical
simulations~\citep{zonta2015growth,yang2018direct}.

While wave growth rates and airflow structure have received much
attention, wind-induced changes to wave shape have been less studied.
Unforced, weakly nonlinear waves on flat bottoms (\eg Stokes, cnoidal,
and solitary waves) are horizontally symmetric about the peak (\ie
zero asymmetry), but are not vertically symmetric (\ie nonzero
skewness, \eg \citep{amick1981periodic,toland2000symmetry}).
Wave shape (skewness and asymmetry) influences many physical phenomena,
such as wave asymmetry sediment
transport~\citep{drake2001discrete,hoefel2003wave} and extreme
waves~\citep{trulsen2012laboratory,trulsen2020extreme}.
Laboratory experiments of wind blowing over periodic waves
demonstrated that wave asymmetry increases with onshore wind speed
in intermediate water~\citep{leykin1995asymmetry} and
deep water~\citep{feddersen2005wind}.
Theoretical studies likewise showed that wind-induced surface
pressure induces wave shape changes in both deep~\citep{zdyrski2020wind}
and shallow~\citep{zdyrski2021wind} water.
However, the influence of wind on wave shape has not yet been
investigated theoretically for shoaling waves on a sloping bottom.

Additionally, the shoaling of unforced waves on bathymetry also induces
wave growth and shape change.
Field observations revealed the importance of nonlinearity in wave
shoaling and its relation to skewness and
asymmetry~\citep{elgar1985observations,freilich1984nonlinear}.
Additionally, laboratory experiments of waves shoaling on planar slopes
yield how the wave height and wave shape evolve with distance up the
beach~\citep{zelt1991run,beji1993experimental,grilli1994shoaling}.
Furthermore, numerical studies investigated wave shoaling all the way to
wave breaking.
A variety of methods were utilized, including pseudospectral
models~\citep{knowles2018shoaling}, the fully nonlinear potential flow
boundary element method
solvers~\citep{grilli1997breaking,derakhti2020unified}, the large eddy
simulation volume of fluid methods~\citep{derakhti2020unified}, and
two-phase direct numerical simulations of both the air and
water~\citep{mostert2020inertial}.
Theoretical~\citep{brun2018convective} and
numerical~\citep{derakhti2020unified} investigations of wave breaking
showed that convective wave breaking depends on the surface water
velocity $u$ and the phase speed $c$ and occurs when the Froude number
$\Fr \coloneqq u/c$ is approximately unity.
The type of wave breaking (\eg spilling, plunging, surging, \etc) is
related to the beach slope $\beta$, initial wave height $H_0$, and
initial wave width $L_0$ through the Iribarren number $\Ir \coloneqq
\beta/\sqrt{H_0/L_0}$~\citep{iribarren1949protection,lara2011reynolds}.

Few studies looked at the combined effects of wind and shoaling of
surface gravity waves.
Experimental studies found that onshore wind increases the surf zone
width~\citep{douglass1990influence} and decreases the wave
height-to-water depth ratio at breaking~\citep{king1996changes}, with
offshore wind having the opposite effect.
Additionally, numerical studies using two-phase RANS solvers of
wind-forced solitary~\citep{xie2014numerical} and
periodic~\citep{xie2017numerical} breaking waves demonstrated that
increasingly onshore winds enhance the wave height at all points prior
to breaking.
Regarding wave shape, only \citet{feddersen2005wind} and
\citet{odea2021field} investigated the combined influence of wind and
shoaling.
\Citet{feddersen2005wind} demonstrated experimentally that onshore winds
enhance the shoaling-induced time-asymmetry.
In field observations of random waves, cross-shore wind was weakly
correlated to the overturn aspect ratio of strongly nonlinear, plunging
waves with offshore (onshore) wind reducing (increasing) the aspect
ratio~\citep{odea2021field}.
However, the wind variation was relatively weak and covariation with
other parameters was not considered.
Despite a growing literature of wave shape measurements and simulations,
a theoretical description of wind-induced changes to wave shoaling
(\eg wave shape, breaking location, \etc) has not yet been
developed.

This study will derive a simplified, theoretical model for wind-forced
shoaling waves that takes the form of a variable-coefficient
Korteweg--de Vries (KdV)--Burgers equation, which is a generalization of
the standard KdV equation.
When the bottom bathymetry is allowed to vary, the coefficients of the
KdV equation are no longer constant and the system is described by a
variable-coefficient KdV (vKdV)
equation~\citep{johnson1973development,svendsen1978deformation}.
The deformation of solitary-wave KdV solutions propagating without wind
on a sloping-bottom vKdV system was studied both
analytically~\citep{miles1979korteweg} and
numerically~\citep{knowles2018shoaling}.
Shoaling causes solitary wave initial conditions to deform and gain a
rear ``shelf'' for small enough slopes~\citep{miles1979korteweg}.
Alternatively, if the flat-bottomed KdV equation is augmented with a
wind-induced surface pressure forcing, the constant-coefficient
KdV--Burgers (KdV--B) equation results~\citep{zdyrski2020wind}.
Wind in the KdV--B equation induces a solitary-wave initial condition to
continuously generate a bound, dispersive, and decaying tail with
polarity depending on the wind direction~\citep{zdyrski2020wind},
analogous to a KdV non-solitary-wave initial
condition~\citep{newell1985solitons}.
The variable-coefficient KdV--Burgers equation combines both shoaling
and wind forcing into one equation.
Our development of a simplified, analytic model for the coupling of
shoaling and wind-forcing highlights the relative importance of these
phenomena and provides a concise framework for analyzing their competing
effects.

In \cref{sec:derivation}, we apply a wind-induced pressure forcing over
a sloping bathymetry to derive a vKdV--Burgers equation and determine a
convective prebreaking condition.
We then solve the resulting vKdV--Burgers equation numerically using a
third-order Runge-Kutta solver and investigate the changes to wave shape
and prebreaking location in \cref{sec:results}.
Finally, we examine the relationship between pressure and wind speed,
isolate the effect of wind from the effect of shoaling, and discuss how
our findings relate to previous laboratory and numerical studies in
\cref{sec:discussion}.

\section{\label{sec:derivation}\NoCaseChange{vKdV}--Burgers equation derivation and
model setup}
\subsection{Governing equations}
\noindent
We derive a vKdV--Burgers equation for wind-forced shoaling waves by
utilizing the standard~\citep{freilich1984nonlinear,grilli2019fully}
simplifying assumptions of planar, two-dimensional waves propagating in
the $+x$-direction on incompressible, irrotational, inviscid fluid
without surface tension.
Additionally, we choose the $+z$-direction to be vertically upwards with
the $z = 0$ at the still water level and impose a bottom bathymetry at
$z = -h(x)$.
The standard incompressibility, bottom boundary, kinematic boundary,
and dynamic boundary conditions are
\begin{alignat}{2}
  0 &= \pdv[2]{\phi}{x} + \pdv[2]{\phi}{z} &&\qq{on}
    -h < z < \eta \,, \label{eq:laplace}\\
  \pdv{\phi}{z} &= - \pdv{h}{x} \pdv{\phi}{x} &&\qq{on} z=-h ,
    \label{eq:bottom_bc}\\
  \pdv{\phi}{z} &= \pdv{\eta}{t} + \pdv{\phi}{x} \pdv{\eta}{x}
    &&\qq{on} z = \eta \,,
  \label{eq:kinematic_bc}\\
  0 &= \frac{p}{\rho_w} + g\eta + \pdv{\phi}{t} +
    \frac{1}{2} \bqty{\pqty{\pdv{\phi}{x}}^2 + \pqty{\pdv{\phi}{z}}^2}
    &&\qq{on} z= \eta \,. \label{eq:dynamic_bc}
\end{alignat}
We introduce the wave profile $\eta(x,t)$, the velocity potential
$\phi(x,z,t)$ derived from the water velocity $\vec{u} = \grad{\phi}$,
the surface pressure $p(x,t)$, the gravitational acceleration $g$, and
the water density $\rho_w$, which is much larger than the air density
$\rho_a \approx \num{1.225e-3} \rho_w$.
Additionally, we remove the Bernoulli constant from the dynamic
boundary condition  by using the $\phi$ gauge freedom.
Next, to examine the wind's effect on shoaling waves, we impose the
analytically simple Jeffreys-type surface pressure $p(x,t)$ forcing
\citep{jeffreys1925formation}
\begin{equation}
  p(x,t) = P \pdv{\eta(x,t)}{x} \,.
  \label{eq:press_def}
\end{equation}
The pressure constant $P \propto \rho_a (U-c)^2$ depends on the wave
phase speed $c$ and wind speed $U$ (\cf{} \cref{sec:press_mag}).
For a wave propagating towards the shore, onshore winds yield $P>0$
whereas offshore winds give $P<0$.

The Jeffreys-type forcing is likely most relevant for near-breaking
waves~\citep{banner1976separation} or strongly forced steep
waves~\citep{touboul2006interaction,tian2013evolution}~(see discussion
in \citet{zdyrski2021wind}).
Recent large eddy simulations~\citep{hao2021mechanistic}, two-phase
direct numerical simulations~\citep{wu2022revisiting}, and
RANS~\citep{xie2017numerical} started investigating the coupling of
periodic waves and wind.
Some simulations~\citep{husain2019boundary,hao2021mechanistic}
suggested a phase shift of approximately \SIrange[range-phrase =
\text{~to~}]{135}{155}{\degree} between waves and the surface
pressure, in contrast to the \SI{90}{\degree} shift predicted by the
Jeffreys-type forcing.
However, these studies used periodic waves and are not
directly applicable to the solitary waves we consider, where an angular
phase shift is undefined.
\Citet{xie2014numerical} considered wind-wave coupling for shoaling
solitary waves, but pressure distributions were not reported.
Therefore, we will prioritize the analytical simplicity of the
Jeffreys-type forcing for our analysis.

\begin{figure}
  \centering
  \includegraphics{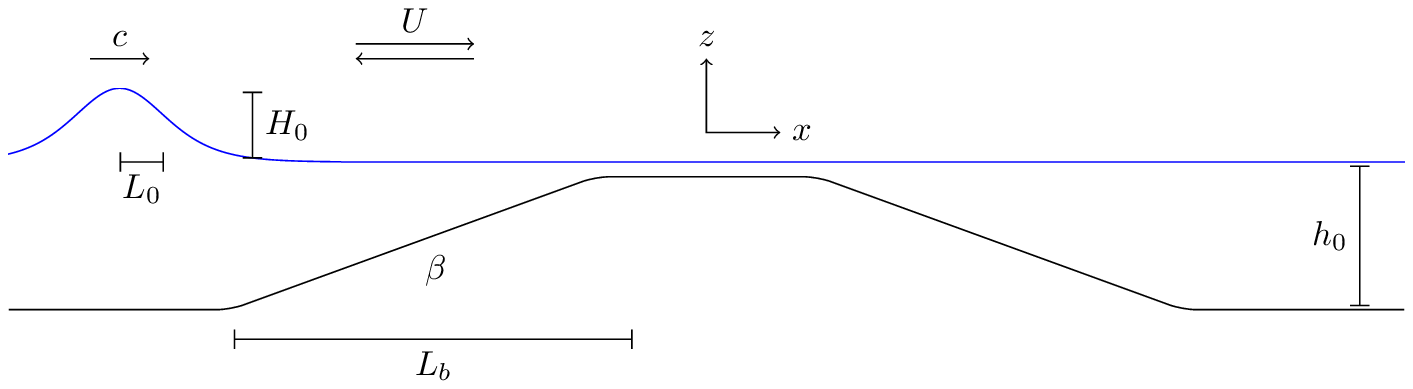}
  \caption{%
    Schematic showing the (periodic) simulation domain and relevant
    lengthscales.
    The blue line represents the water surface and wave profile $\eta$
    and the solid black line is the bottom bathymetry $h(x)$.
    The solitary wave initial condition has a height $H_0$ and effective
    half-width $L_0$ and begins with its peak on the far left
    side, in the middle of flat region of depth $h_0$.
    The initial wave then propagates to the right with phase speed $c$
    up the beach with slope $\beta$ until it reaches prebreaking (\cf{}
    \cref{sec:breaking_criteria}).
    The positive/negative wind speed $U$ corresponds to an
    onshore/offshore wind forcing.
  }\label{fig:schematic}
\end{figure}

\subsection{\label{sec:domain}Model domain and model parameters}
The model domain (\cref{fig:schematic}) is similar to that of
\citet{knowles2018shoaling} and consists of an initial flat section of
length $L_f$ at a depth of $h_0$ and transitions smoothly at $x=0$ into
a planar beach region with constant slope $\beta$ and characteristic
beach width $L_b \coloneqq h_0/\beta$.
The bathymetry then smoothly transitions to a flat plateau of length $2
L_f$ at a depth of $h = 0.1 h_0$ followed by a downward slope with
slope $-\beta$.
Finally, there is another flat section of length $L_f$ at a depth of
$h_0$ before the domain wraps periodically, which simplifies the
boundary conditions.

The initial condition will be a KdV solitary wave with height $H_0$ and
width $L_0$ following \citet{knowles2018shoaling}, and $L_0$ will be
specified later.
The solitary wave begins centered on the left boundary, in between the
two flat, deep sections of length $L_f$.
From the defined dimensional quantities, we specify four nondimensional
parameters
\refstepcounter{equation}
\begin{equation}
  \epsilon_0 \coloneqq \frac{H_0}{h_0}, \qquad
  \mu_0 \coloneqq \pqty{\frac{h_0}{L_0}}^2, \qquad
  P_0 \coloneqq \frac{P}{\rho_w g L_0} \,, \qquad
  \gamma_0 \coloneqq \frac{L_0}{L_b} \,.
  \tag{\theequation\textit{a--d}}
\end{equation}
Here, $\epsilon_0$ is the nondimensional initial wave height, $\mu_0$ is
the square of the nondimensional initial inverse wave width, $P_0$ is the
nondimensional pressure magnitude (normalized by the initial wave
width), and $\gamma_0$ is ratio of the initial wave width to the beach
width.
Note that the wave width-to-beach width ratio $\gamma_0$ is related to the
beach slope $\beta$ as $\gamma_0 = \beta/\sqrt{\mu_0}$.
Together, these four nondimensional parameters control the system's
dynamics.

\subsection{Nondimensionalization}
We nondimensionalize our system's variables using the characteristic
scales described in \cref{sec:domain}:
the initial depth $h_0$;
the initial wave's height $H_0$;
the initial wave's horizontal lengthscale $L_0$;
the gravitational acceleration $g$;
and the pressure magnitude $P$.
Using primes for nondimensional variables, we normalize as
\citet{zdyrski2021wind} did and define
\begin{equation}
  \begin{aligned}[t]
  x &= L_0 x' = h_0 \frac{x'}{\sqrt{\mu_0}}\,, \\
  z &= h_0 z' \,, \\
  t &= \frac{t' L_0}{\sqrt{g h_0}}
    = \frac{t'}{\sqrt{\mu_0}} \sqrt{\frac{h_0}{g}} \,,
  \end{aligned}
  \qquad
  \begin{aligned}[t]
  h &= h' h_0 \,, \\
  \eta &= H_0 \eta' = h_0 \epsilon_0 \eta' \,, \\
  \phi &= \phi' H_0 L_0 \sqrt{\frac{g}{h_0}} =
    \frac{\phi'\epsilon_0}{\sqrt{\mu_0}}\sqrt{g h_0^3} \,. \\
  \end{aligned}
  \label{eq:nondim_expressions}
\end{equation}
We later assume the nondimensional parameters $\epsilon_0$, $\mu_0$,
$\gamma_0$, and $P_0$ are small to leverage a perturbative analysis.
For the constant slope $\beta$ beach profile, the spatial derivative of
the bathymetry is also small $\partial_{x'} h' = \beta/\sqrt{\mu_0} =
\gamma_0 \ll 1$ (the factor of $\sqrt{\mu_0}$ comes from the different
nondimensionalizations of $h$ and $x$).
However, perturbation analyses are simplest when all nondimensional
variables are $\order{1}$.
Therefore, we leverage the two, horizontal lengthscales $L_0$ and $L_b$
(\cf{} \cref{sec:domain}) to define a nondimensional, stretched
bathymetry $\htil'$ that depends on $x/L_b = \gamma_0 x'$ as
$\htil'(\gamma_0 x') = h'(x')$.
Then, denoting derivatives with respect to $\gamma_0 x'$ using an overdot,
the derivative of $\htil$ is $\dhtil' \coloneqq \partial_{\gamma_0 x'}
\htil'(\gamma_0 x') = \order{1}$, and the small slope becomes explicit as
$\partial_{x'} h' = \gamma_0 \dhtil'$.

Now, the nondimensional equations take the form
\begin{alignat}{2}
  0 &= \mu_0 \pdv[2]{\phi'}{{x'}} + \pdv[2]{\phi'}{{z'}} &&\qq{on}
    -1 < z' < \epsilon_0 \eta' \,, \label{eq:laplace_nondim} \\
    \pdv{\phi'}{z'} &= -\mu_0 \gamma_0 \dhtil' \pdv{\phi'}{x'}
    &&\qq{on} z'=-\htil'(\gamma_0 x') \,, \label{eq:bottom_bc_nondim} \\
  \pdv{\phi'}{z'} &= \mu_0 \pdv{\eta'}{t'} + \epsilon_0 \mu_0
    \pdv{\phi'}{x'} \pdv{\eta'}{x'} &&\qq{on} z' = \epsilon_0 \eta' \,,
    \label{eq:kinematic_bc_nondim} \\
  0 &= P_0 \pdv{\eta'}{x'} + \eta' + \pdv{\phi'}{t'} + \frac{1}{2}
    \bqty{\epsilon_0 \pqty{\pdv{\phi'}{x'}}^2
    + \frac{\epsilon_0}{\mu_0} \pqty{\pdv{\phi'}{z'}}^2}
    &&\qq{on} z'= \epsilon_0 \eta' \,.
    \label{eq:dynamic_bc_nondim}
\end{alignat}
For the remainder of \cref{sec:derivation}, we remove the primes for
clarity.

\subsection{\label{sec:boussinesq}Boussinesq equations, multiple-scale
expansion, and vKdV--Burgers equation}
\noindent
We follow the conventional Boussinesq equation derivation presented in,
\eg, \citet{mei2005nonlinear} or \citet{ablowitz2011nonlinear}.
The two modifications we include are the weakly sloping bottom, similar
to the treatment in \citet{johnson1973development} and
\citet{mei2005nonlinear}, and the inclusion of a pressure forcing like
that of \citet{zdyrski2021wind}.
However, the joint contributions of both pressure and shoaling are new.
For the sake of brevity, we only detail the relevant differences here,
but we still treat the derivation formally to ensure a proper ordering
of small terms and obtain the parameters' validity ranges.
First, we expand the velocity potential in a Taylor series about the
bottom $z=-h(x)$ as
\begin{equation}
  \phi(x,z,t) = \sum_{n=0}^\infty \bqty{z+\htil(\gamma_0 x)}^n\phi_n(x,t)
  \,.
\end{equation}
Substituting this expansion into the incompressibility
\cref{eq:laplace_nondim} and bottom boundary condition
\labelcref{eq:bottom_bc_nondim} and assuming $\mu_0 \ll 1$ gives $\phi$ as a
function of the velocity potential evaluated at the bottom $\varphi
\coloneqq \phi_0$.
If we further assume that the bottom is very weakly sloping
$\gamma_0 \sim \mu_0 \ll 1$, this simplifies to
\begin{equation}
    \phi = \varphi
    - \mu_0 \frac{1}{2} (z+\htil)^2 \partial_x^2 \varphi
    + \order{\mu_0^2, \gamma_0^2, \gamma_0 \mu_0} \,.
    \label{eq:phi_expansion}
\end{equation}
Note that the assumption $\gamma_0 \sim \mu_0 \ll 1$ implies a moderate
slope $\beta = \gamma_0 \sqrt{\mu_0} \sim \mu_0^{3/2}$ and is used by
several other
authors~\citep{johnson1973development,miles1979korteweg,knowles2018shoaling}.
For reference, if $\mu_0 = \gamma_0 = 0.1$, then this implies a
physically realistic $\beta = 0.03$.

Substituting this $\phi$ expansion \labelcref{eq:phi_expansion} into the
kinematic boundary condition \labelcref{eq:kinematic_bc_nondim} and dynamic
boundary condition \labelcref{eq:dynamic_bc_nondim} yields Boussinesq-type
equations with a pressure-forcing term
\begin{gather}
  \partial_t \eta + \pqty{\htil + \epsilon_0 \eta} \partial^2_x \varphi \
    + \pqty{\gamma_0 \dhtil + \epsilon_0 \partial_x \eta} \partial_x \varphi
    - \mu_0 \frac{1}{6} \htil^3 \partial^4_x \varphi
    = \order{\mu_0^2, \gamma_0^2, \gamma_0 \mu_0}
    \label{eq:kinematic_bc_varphi} \,, \\
  P_0 \partial_x \eta + \eta + \partial_t \varphi
    - \frac{1}{2} \mu_0 \htil^2 \partial^2_x \partial_t \varphi
    + \frac{1}{2} \epsilon_0 \pqty{\partial_x \varphi}^2
    = \order{\mu_0^2, \gamma_0^2, \gamma_0 \mu_0} \,.
    \label{eq:dynamic_bc_varphi}
\end{gather}
Note that replacing $\htil$ with the total depth $h_{\text{total}} =
\htil + \epsilon_0 \eta$ shows that these are equivalent to the
flat-bottomed Boussinesq equations with $h_{\text{total}} = 1 +
\epsilon_0 \eta$.
In other words, any sloping-bottom terms $\dhtil$ only appear in the
combination $\partial_x h_{\text{total}} = \gamma_0 \dhtil + \epsilon_0
\partial_x \eta$.
This is expected since the only sloping-bottom term $\mu_0 \gamma_0 \dhtil
\partial_x \phi$ in the governing equations
\labelcref{eq:laplace_nondim,eq:bottom_bc_nondim,eq:kinematic_bc_nondim,eq:dynamic_bc_nondim}
was dropped when we neglected terms of $\order{\mu_0^2, \gamma_0^2, \gamma_0
\mu_0}$.

Since the bathymetry varies on the slow scale $x/\gamma_0$, we expand
our system in multiple spatial scales $x_n = \gamma_0^n x$ for $n=
0,1,2,\ldots$, so the derivatives become
\begin{equation}
  \pdv{x} \to \pdv{x_0} + \gamma_0 \pdv{x_1} + \ldots \,,
\end{equation}
and the bathymetry is a function of the long spatial scale $\htil =
\htil(x_1)$.
Then, we expand $\eta$ and $\varphi$ in asymptotic series of $\epsilon_0$
\refstepcounter{equation}
\begin{equation}
  \eta(x,t) \to \sum_{k=0}^{\infty} \epsilon_0^k
    \eta_{k}(t, x_0,x_1,\ldots) \,, \qquad
  \varphi(x,t) \to \sum_{k=0}^{\infty} \epsilon_0^k
    \varphi_{k}(t, x_0,x_1,\ldots) \,.
  \tag{\theequation\textit{a,b}}
\end{equation}
Similar to \citet{johnson1973development}, we replace $x_0$ and $t$ with
left- and right-moving coordinates translating with speed $\ctil(x_1)$
dependent on the stretched coordinate $x_1$:
\begin{equation}
  \xi_+ = - t + \int^{x_0} \frac{\dd{x'_0}}{\ctil(\gamma_0 x'_0)} \,,
  \qquad
  \xi_- = t + \int^{x_0} \frac{\dd{x'_0}}{\ctil(\gamma_0 x'_0)} \,.
\end{equation}
Then, we replace the derivatives $\partial_t$ and $\partial_{x_0}$ with
\begin{equation}
  \pdv{t} = \pdv{\xi_-} - \pdv{\xi_+} \,, \qquad
  \pdv{x_0} = \frac{1}{\ctil} \pqty{ \pdv{\xi_-} + \pdv{\xi_+} } \,.
\end{equation}
Now, we will assume that $\epsilon_0 \sim P_0 \sim \mu_0 \ll 1$ and
follow the standard multiple-scale
technique~\citep{mei2005nonlinear,ablowitz2011nonlinear}.
The order-one terms $\order{\epsilon_0^0}$ from
\cref{eq:kinematic_bc_varphi,eq:dynamic_bc_varphi} yield wave
equations for $\phi_0$ and $\eta_0$
\refstepcounter{equation} \label{eq:order1-wave}
\begin{equation}
  \pdv{\phi_0}{\xi_+}{\xi_-}  = 0 \,, \qquad
  \pdv{\eta_0}{\xi_+}{\xi_-}  = 0 \,,
  \tag{\theequation\textit{a,b}}
\end{equation}
with right-moving solutions
\begin{equation}
  \varphi_0 = f_0(\xi_+,x_1) \qq{and}
  \eta_0 = \partial_{\xi_+} f_0(\xi_+,x_1) \,,
\end{equation}
propagating with the slowly varying, linear shallow-water phase speed
$\ctil(x_1) = \sqrt{\htil(x_1)}$.
Continuing to $\order{\epsilon_0}$ of the asymptotic expansion gives
\begin{gather}
  \begin{split}
      &- \pdv{\eta_1}{\xi_+} + \pdv{\eta_1}{\xi_-}
      + \pdv[2]{\varphi_1}{\xi_+}
      + 2 \pdv{\varphi_1}{\xi_+}{\xi_-}
      +  \pdv[2]{\varphi_1}{\xi_-}
    =
      - 2 \frac{\gamma_0}{\epsilon_0} \ctil \pdv{\varphi_0}{\xi_+}{x_1}
      + \frac{\gamma_0}{\epsilon_0} \pdv{\ctil}{x_1} \pdv{\varphi_0}{\xi_+}
    \\ &\qquad
      - \frac{1}{\ctil^2} \eta_0 \pdv[2]{\varphi_0}{\xi_+}
      - \frac{\gamma_0}{\epsilon_0} \frac{\dhtil}{\ctil} \pdv{\varphi_0}{\xi_+}
      - \frac{1}{\ctil^2} \pdv{\eta_0}{\xi_+} \pdv{\varphi_0}{\xi_+}
      + \frac{\mu_0}{\epsilon_0} \htil \frac{1}{6} \pdv[4]{\varphi_0}{\xi_+}
      \,,
  \end{split}
    \\
  \begin{split}
    &\eta_1 - \pdv{\varphi_1}{\xi_+} + \pdv{\varphi_1}{\xi_-} =
    - \frac{P_0}{\epsilon_0} \frac{1}{\ctil} \pdv{\eta_0}{\xi_+}
    - \frac{1}{2} \frac{\mu_0}{\epsilon_0} \htil \frac{\partial^3 \varphi_0}{\partial^3 \xi_+}
    - \frac{1}{2 \ctil^2} \pqty{\pdv{\varphi_0}{\xi_+}}^2
    \,.
  \end{split}
\end{gather}
Eliminating $\eta_1$ from these equations gives
\begin{equation}
  4 \pdv{\phi_1}{\xi_+}{\xi_-} =
  - 2\frac{\gamma_0}{\epsilon_0} \ctil \pdv{\eta_0}{x_1}
  - \frac{\gamma_0}{\epsilon_0} \pdv{\ctil}{x_1} \eta_0
  - 3 \frac{1}{\ctil^2} \eta_0 \pdv{\eta_0}{\xi_+}
  - \frac{1}{3} \frac{\mu_0}{\epsilon_0} \ctil^2 \pdv[3]{\eta_0}{\xi_+}
  - \frac{1}{\ctil} \frac{P_0}{\epsilon_0} \pdv[2]{\eta_0}{\xi_+}
  \,.
\end{equation}
The left-hand operator $\partial^2/\partial \xi_- \partial \xi_+$ is the
same as the $\order{1}$ differential operator \labelcref{eq:order1-wave}.
Therefore, the right-hand side must vanish to prevent $\phi_1$ from
developing secular terms.
Thus, the right-hand side becomes the variable-coefficient Korteweg--de
Vries--Burgers (vKdV--Burgers) equation
\begin{equation}
  \frac{\gamma_0}{\epsilon_0} \ctil \pdv{\eta_0}{x_1}
  + \frac{1}{2} \frac{\gamma_0}{\epsilon_0} \pdv{\ctil}{x_1} \eta_0
  + \frac{3}{2} \frac{1}{\ctil^2} \eta_0 \pdv{\eta_0}{\xi_+}
  + \frac{1}{6} \frac{\mu_0}{\epsilon_0} \ctil^2 \pdv[3]{\eta_0}{\xi_+}
  + \frac{1}{2 \ctil} \frac{P_0}{\epsilon_0} \pdv[2]{\eta_0}{\xi_+}
  = 0 \,.
  \label{eq:vckdv_burgers}
\end{equation}
Finally, multiplying \cref{eq:vckdv_burgers} by $\epsilon_0$, adding the
$\order{1}$ differential equation $\partial_{\xi_-} \eta_0 = 0$ derived
from \cref{eq:order1-wave}, and transforming back to the original,
nondimensional variables $x$ and $t$ yields
\begin{equation}
  \pdv{\eta_0}{t}
  + c \pdv{\eta_0}{x}
  + \frac{1}{2} \pdv{c}{x} \eta_0
  + \frac{3}{2} \epsilon_0 \frac{1}{c} \eta_0 \pdv{\eta_0}{x}
  + \frac{1}{6} \mu_0 c^5 \pdv[3]{\eta_0}{x}
  + \frac{1}{2} P_0 c \pdv[2]{\eta_0}{x}
  = 0 \,.
  \label{eq:vckdvb_lab_frame}
\end{equation}
This vKdV--Burgers equation represents an analytically simple system for
studying the effects of wind-forcing and shoaling.
In the absence of wind-forcing ($P_0=0$), it reduces to the
variable-coefficient KdV equation~\citep{newell1985solitons}, while it
simplifies to the constant-coefficient KdV--Burgers equation in the case
of flat bathymetry ($c=1$)~\citep{zdyrski2021wind}.
The formal derivation revealed the depth-dependent $c$ factor in the
wind-forced $P_0$ term which was absent in the flat-bottomed case.
The pressure term $P_0 \partial^2_x \eta_0$ functions as a damping,
positive viscosity for offshore $P_0<0$ wind, making
\cref{eq:vckdvb_lab_frame} a (forward) vKdV--Burgers equation.
Conversely, onshore $P_0>0$ wind causes a growth-inducing, negative
viscosity giving the backward vKdV--Burgers equation.
Though the backward, constant-coefficient KdV--Burgers equation is ill
posed in the sense of Hadamard~\citep{hadamard1902problemes}, this is
irrelevant here owing to the finite time the wave takes to reach the
beach.

\subsection{\label{sec:ICs}Initial conditions}
Our initial condition will be the solitary-wave solutions of the
unforced ($P_0=0$), flat-bottom KdV equation.
These waves balance the KdV equation's nonlinearity $\eta_0 \partial_x
\eta_0$ and dispersion $\partial_x^3 \eta_0$ terms, propagate without
changing shape, and require that the height $H_0$ and width $L_0$ satisfy
the constraint $H_0 L_0^2 = \text{constant}$.
Therefore, we now fix the previously unspecified $L_0$ by choosing
$\mu_0 = (3/4) \epsilon_0$ so $L_0$ acts like an effective
half-width for the solitary wave initial
condition~\citep{mei2005nonlinear}
\begin{equation}
  \eta_0 = \sech^2\pqty{x} \,.
  \label{eq:initial_condition}
\end{equation}
While the unforced KdV equation also possesses periodic solutions called
cnoidal waves, we only consider solitary waves here.

\subsection{\label{sec:breaking_criteria}Convective breaking criterion}
The asymptotic assumptions used to derive the vKdV--Burgers equation
\labelcref{eq:vckdvb_lab_frame} fail when the wave gets too large.
Therefore, we require a condition to determine when the simulations
should stop.
\Citet{brun2018convective} defined a convective breaking condition for
solitary waves on a flat-bottom depending on the surface water velocity
$u_s(x,t)$ and the phase speed $c$ using the local Froude number $\Fr
\coloneqq \epsilon_0 u_s(x,t)/c(t)$, with the $\epsilon_0$ coming from
nondimensionalization.
Convective breaking occurs wherever $\max_x(\Fr) = 1$, where $\max_x$
represents the maximum over $x$.
However, when the Froude number approaches the breaking value of unity,
our weakly nonlinear asymptotic assumption used to derive the
vKdV--Burgers equation is violated.
Thus, we instead stop our simulations at the smaller prebreaking Froude
number $\Frpb \coloneqq 1/3$ and define the ``prebreaking'' time $\tpb$
as the first time such that $\max_x (\Fr) = \Frpb \coloneqq 1/3$.
Likewise, we define $\xpb$ as the location where $\Fr = \Frpb$, which
will be very near the wave peak.

As the solitary wave propagates on a slope, the wave evolves
over time and the phase speed $c$ can be ambiguous.
One option is to use the adiabatic approximation derived by the authors
of \citep{miles1979korteweg} for unforced solitary waves on very gentle
slopes
\begin{equation}
  \cadi(t) = \sqrt{h[\xpeak(t)]}
    \pqty{1 + \frac{\epsilon_0}{2} \frac{\eta[\xpeak(t)]}{h[\xpeak(t)]}}
    \,,
  \label{eq:c_adiabatic}
\end{equation}
with $\xpeak$ the location of the wave peak.
Alternatively, \citet{derakhti2020unified} used large eddy simulations
to numerically investigate unforced solitary wave breaking on slopes
ranging from $\beta = \numrange{0.2}{0.005}$ for two different forms of
$c$.
They found wave breaking at $\max_x(\Fr) = 0.85$ when using the speed of
the numerically tracked wave peak $\cpeak$.
However, they also found that the shallow-water approximation
$c_{\text{shallow}} = \sqrt{h(\xpeak) + \epsilon_0
\eta(\xpeak)}$ [equivalent to $\cadi$ to
$\order{\epsilon_0^2}$] was within \SI{15}{\percent} of $\cpeak$
near breaking.
Therefore, we will use \cref{eq:c_adiabatic} for $c(t)$ owing to its
simplicity and theoretical foundation.
Finally, though these studies all considered unforced solitary waves,
our results will show that $\cadi$ varies approximately \SI{3}{\percent}
across pressure magnitudes $P_0$ for our simulations, so this
approximation is valid.

We still need an expression for the wave velocity at the surface
$u_s(x,t)$, which we now derive by modifying the example of
\citet{brun2018convective} to include sloping bathymetry and pressure
forcing.
Combining the vKdV--Burgers equation \labelcref{eq:vckdvb_lab_frame} and
kinematic boundary condition \labelcref{eq:kinematic_bc_varphi} eliminates
$\partial_t \eta$ yielding
\begin{equation}
  \begin{split}
  & \ctil^2 \pdv[2]{\varphi}{x}
  - \ctil \pdv{\eta}{x}
  + \epsilon_0 \pqty{
    \eta \pdv[2]{\varphi}{x}
    + \pdv{\eta}{x} \pdv{\varphi}{x}
    - \frac{3}{2} \frac{1}{\ctil} \eta \pdv{\eta}{x}
  }
  - P_0 \frac{1}{2} \ctil \pdv[2]{\eta}{x}
  \\ & \qquad
  - \mu_0 \pqty{
    \frac{1}{6} \ctil^6 \pdv[4]{\varphi}{x}
    + \frac{1}{6} \ctil^5 \pdv[3]{\eta}{x}
  }
  + \gamma_0 \pqty{
    2 \ctil \dctil \pdv{\varphi}{x}
    - \frac{1}{2} \eta \dctil
  }
  = 0 \,.
  \end{split}
  \label{eq:vckdvb_and_kinematic}
\end{equation}
Here, $\dctil \coloneqq \partial_{x_1} \ctil(x_1) = \order{1}$ in
analogy to the previously defined $\dhtil$.
Assuming an ansatz
\begin{align}
    \pdv{\varphi}{x} &= \frac{1}{\ctil} \eta + \epsilon_0 A(x,t) + \gamma_0
      B(x,t) + \mu_0 C(x,t) + P_0 D(x,t)
  \label{eq:varphi_expr_dx1}
      \\
    \implies
    \pdv[2]{\varphi}{x} &= \frac{1}{\ctil} \pdv{\eta}{x} + \epsilon_0
    \pdv{A}{x}
    + \gamma_0 \pqty{\pdv{B}{x} - \frac{1}{\ctil^2} \eta \dctil}
    + \mu_0 \pdv{C}{x} + P_0 \pdv{D}{x} \,,
  \label{eq:varphi_expr_dx2}
\end{align}
we insert \cref{eq:varphi_expr_dx1,eq:varphi_expr_dx2} into
\cref{eq:vckdvb_and_kinematic}, drop terms of $\order{\epsilon_0^2}$ and
solve for $A$, $B$, $C$, and $D$ by using the independence of
$\epsilon_0$, $\gamma_0$, $\mu_0$, and $P_0$:
\refstepcounter{equation} \label{eq:froude_terms}
\begin{equation}
  A = -\frac{1}{4 \ctil^3} \eta^2
  \,, \qquad
  B = -\frac{1}{2 \ctil^2} \int^x_{-\infty} \eta(x') \dctil(\gamma_0 x') \dd{x'}
  \,, \qquad
  C = \frac{\ctil^3}{3} \pdv[2]{\eta}{x}
  \,, \qquad
  D = \frac{1}{2\ctil} \pdv{\eta}{x}
  \,.
  \tag{\theequation\textit{a--d}}
\end{equation}
Note that $A$ represents the nonlinear contribution, $B$ the effect of
shoaling, $C$ the dispersive effect, and $D$ the pressure forcing.
Finally, the Taylor expansion of $\phi(x,z)$ \cref{eq:phi_expansion}
gives the fluid velocity at the surface $u_s(x,t)$ using $u=\partial_x
\phi$ evaluated at $z=\epsilon_0 \eta$:
\begin{equation}
  \begin{split}
    u_s(x,t) &= \partial_x \varphi - \mu_0 \frac{1}{2} \ctil^4 \partial^3_x \varphi
    \\ &=
    \frac{1}{\ctil} \eta
    - \epsilon_0 \frac{1}{4 \ctil^3} \eta^2
    + P_0 \frac{1}{2 \ctil} \pdv{\eta}{x}
    - \mu_0 \frac{\ctil^3}{6} \pdv[2]{\eta}{x}
    - \gamma_0 \frac{1}{2 \ctil^2} \int_{-\infty}^x \eta(x') \dctil(\gamma_0 x') \dd{x'}
    \,.
  \end{split}
  \label{eq:water_velocity}
\end{equation}
Therefore, the Froude number is calculated as
\begin{equation}
  \Fr \coloneqq \frac{\epsilon_0 u_s(\xpeak,\tpeak)}{\sqrt{h(\xpeak)}}
  \pqty{1 + \frac{\epsilon_0}{2} \frac{\eta(\xpeak)}{h(\xpeak)}}^{-1}
  \,,
  \label{eq:froude_num}
\end{equation}
with $u_s(x,t)$ given by \cref{eq:water_velocity}.
\Cref{eq:froude_num} demonstrates that the choice of $\Frpb=1/3$ keeps
the problem in the correct asymptotic regime.
Our simulated waves will reach prebreaking in depths of
$h(\xpb)=\numrange{0.59}{0.72}$, so
\cref{eq:water_velocity,eq:froude_num} show that $\Frpb = 1/3$
corresponds to a dimensional $H(\xpb)/h(\xpb) \approx \Frpb/\sqrt{h(\xpb)} =
\numrange{0.39}{0.43}$.
Considering an asymptotic regime around $\epsilon_0$ of
$\epsilon_0^{3/2}$ to $\epsilon_0^{1/2}$, or \numrange{0.089}{0.45}, we
see that $H(x)/h(x)$ remains in the asymptotic regime.%

\subsection{Numerics}
\begin{table}
  \centering
  \caption{%
    Range of nondimensional parameters simulated.
  }\label{tab:parameters}
  \begin{tabular*}{0.75\linewidth}{ l@{\extracolsep{\fill}} c }
    \toprule \toprule
    Parameter & Range \\
    \midrule
    $\epsilon_0$ & 0.2 \\
    $\mu_0 $ & 0.15 \\
    $\abs{P / (\rho_w g L_0 \epsilon_0)}$ & 0, 0.00625, 0.0125, 0.025, 0.05 \\
    $\beta$ & 0.01, 0.015, 0.02, 0.025 \\
    \bottomrule \bottomrule
  \end{tabular*}
\end{table}
The vKdV--Burgers equation \labelcref{eq:vckdvb_lab_frame} lacks
analytic, solitary-wave-type solutions, so we solve it numerically using
a third-order explicit Runge-Kutta adaptive time stepper with the error
controlled by a second-order Runge-Kutta method as implemented in
\textsc{SciPy}~\citep{virtanen2020scipy}.
The domain's deep, flat regions have length $L_f = 20 L_0$.
Therefore, the spatial domain's entire, periodic width $L_x$ varies
between \num{108} and \num{150}, depending on the slope $\beta$.
We discretize the spatial domain using a fourth-order finite difference
method with $N_x = 2^2 \roundeven{L_x/0.025/2^2}$ points which yields a
grid spacing of $\dd{x} = L_x/N_x \approx 0.025$ and ensures that the
grid can be refined by two steps.
We employ adaptive time stepping to keep the relative error below
\num{e-3} and the absolute error below \num{e-6} at each step.
For all cases, the average time step is $\Delta t \approx \num{2e-4}$.
The pressure is initially turned off until the solitary wave is one unit
(\ie a half-width $L_0$) away from the start of the beach slope.
The pressure is linearly ramped up to its full value over the time the
wave takes to cross a full-width $2 L_0$.
We included a hyperviscosity $\nu_4 \partial_x^4 \eta_0$ with $\nu_4 =
\num{1e-5}$ for numerical stability~\citep{cho1996emergence}.
Finally, we only considered solitary waves, as mentioned previously,
because cnoidal wave trains require spin-up and damping since we only
consider prebreaking waves.

We validated the solver against the unforced, flat-bottom analytical
solution and had a normalized root-mean-square error of \num{1.6e-4}
after nondimensional time $t = 50$ (the longest simulation time)
as well as a normalized wave height change of $1-\bqty{\max(\eta_0) -
\min(\eta_0)}/\allowbreak \bqty{\max(\eta^{(0)}_0) - \min(\eta^{(0)}_0)}
= \num{1.4e-4}$.
To verify the numerical convergence on our nominal (fine) grid size, we
also ran the simulation on a medium grid and a fine grid, each with a
refinement ratio of \num{2}.
We then calculated the Romberg interpolation~\citep{roy2003grid} and
compared each grid's normalized root-mean-square error with respect to
this interpolation to yield a grid refinement index
($GCI$)~\citep{roache1998verification} for the unforced case with the
steepest $\beta = 0.025$ and shallowest $\beta = 0.01$ slopes.
Both slopes had a convergence order $p > \num{2.1}$ which yielded grid
convergence indices indices $GCI_{\text{nominal,medium}} < \num{3e-4}$
and $GCI_{\text{medium,coarse}} < \num{8e-4}$ much less than unity and
implying that the results are grid-converged.
Furthermore, the profiles for all times until prebreaking were nearly
identical to the simulations of \citet{knowles2018shoaling} for an
unforced solitary wave shoaling on a slope.
Finally, the simulation reproduced the finding of
\citet{knowles2018shoaling} that small waves ($\epsilon_0 \ll 1$) on weak
slopes ($\gamma_0 \ll 1$) yield Green's Law for the wave height $H(x)
\coloneqq \max_t (\eta) \propto h(x)^{1/4}$ (with $\max_t$ the maximum
over time $t$), while moderate waves ($\epsilon_0 < 1$) on very weak
slopes ($\gamma_0 \lll 1$) give Miles' adiabatic law $H(x) \propto
h(x)^{-1}$~\citep{miles1983solitary}.

The vKdV--Burgers equation \labelcref{eq:vckdv_burgers} is determined
by two nondimensional parameter combinations: the pressure term
$P_0/\epsilon_0$ and the shoaling term $\gamma_0/\epsilon_0$.
Recall that the dispersive term $\mu_0/\epsilon_0$ is a redundancy which
we fixed by specifying $L_0$ (\cf{} \cref{sec:ICs}).
We investigate this two-dimensional parameter space by choosing
$\epsilon_0 = 0.2$ and $\mu_0 = 0.15$ and varying the beach slope $\beta
= \numrange{0.01}{0.025}$ and pressure $\abs{P} = \numrange[range-phrase
= \text{~to~}]{0}{0.05}$
(\cf{} \cref{sec:press_mag} for a discussion of the size of $P$).
This yields a total of \num{36} simulations (\cref{tab:parameters}).
Note that \cref{eq:vckdv_burgers} demonstrates changing $\epsilon_0 \to
\lambda \epsilon_0$ is equivalent to $\gamma_0 \to \gamma_0/\lambda$ in
the wave's comoving reference frame.
Therefore, solutions for waves with different initial heights
$\epsilon_0$ can be generated from our solutions to the vKdV--Burgers
equation in the laboratory frame \cref{eq:vckdvb_lab_frame} by scaling
the height, boosting, and adjusting $\gamma_0$.
Note the asymptotic expansion assumed $P_0 \sim \epsilon_0$, or
$P/(\rho_w g L_0 \epsilon_0) \sim 1$, but the pressure values we are
using (\cref{tab:parameters}) are smaller than unity.
Nevertheless, multiple-scale expansions are often accurate outside their
parameters' validity ranges and this constraint would be satisfied
asymptotically for smaller values of $\epsilon_0$.

\subsection{\label{sec:shape_params}Shape statistics}
\begin{figure}
  \centering
  \includegraphics{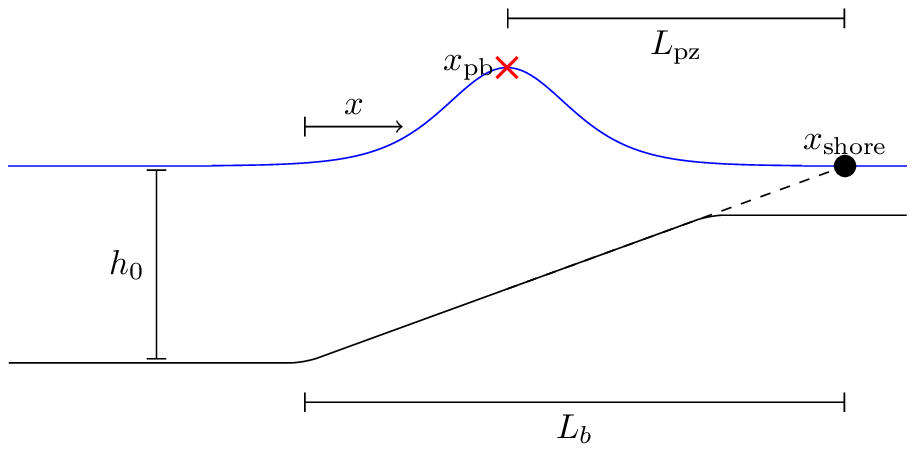}
  \caption{%
    Schematic showing the definition of the prebreaking zone and shore
    locations.
    The blue line represents the water surface and wave profile $\eta$
    at prebreaking and the solid black line is the bottom bathymetry
    $h(x)$.
    The bathymetry consists of a flat region of depth $h_0$, a sloping
    region, and a shallow plateau.
    The shoreline $x_{\text{shore}}$ (black dot) is the location where
    the bathymetry would intersect the still water level if it had a
    constant slope (dashed line).
    The beach width $L_b$ is the distance from the toe of the beach
    slope to $x_{\text{shore}}$ and the prebreaking point $\xpb$ is
    the location on the wave where $\Fr = \Frpb$, which will be very
    near the wave peak.
    The prebreaking zone width $\Lpz$ is the distance from $\xpb$ to
    $x_{\text{shore}}$.
  }\label{fig:break-point-schematic}
\end{figure}

While previous studies investigated shoaling's effect on wave-breaking
location, we will theoretically examine the combined influences of both
wind and shoaling on wave location $\xpb$ at prebreaking $\tpb$
(\cref{sec:breaking_criteria}).
To estimate how $\xpb$ changes, we first calculate the shoreline
$x_{\text{shore}}$ as the location where the bathymetry would intersect
$z=0$ if it had a constant slope $\beta$ without our shallow plateau
(\cref{fig:break-point-schematic}).
Then, we calculate the prebreaking zone width as $\Lpz \coloneqq
\xpb - x_{\text{shore}}$.
For a given beach slope $\beta$, we will analyze the change in
the prebreaking zone width relative to the unforced case $\Delta \Lpz
\coloneqq \Lpz - \eval{\Lpz}_{P=0}$ normalized by the unforced
prebreaking zone width $\eval{\Lpz}_{P=0}$.
This bulk statistic $\Delta \Lpz/(\eval{\Lpz}_{P=0})$ determines the
variance in prebreaking locations as a fractional change of the
prebreaking zone width.

Additionally, since we wish to analyze the effect of wind and shoaling
on wave shape, we will investigate four shape statistics that vary as
the wave propagates.
The first three are local shape parameters defined at each location $x$.
First, we directly examine the maximum Froude number $\max_t(\Fr)$
expressed in \cref{eq:froude_num}.
Second, we investigate the maximum height relative to the local water
depth $\max_t(\eta)/h(x)$ at each location $x$.
Third, we consider the maximum slope $\max_t(\abs{\partial \eta /
\partial x})$.
Both the relative height and maximum slope contribute to the convective
breaking criterion $\max_x (\Fr) = \Frpb$.
Finally, we introduce a bulk shape parameter, the full width of the
wave at half of the wave's maximum (FWHM) $L_W(t)$ normalized by the
local water depth $h(x)$.
For our unforced KdV solitary wave initial condition
\labelcref{eq:initial_condition}, the FWHM divided by the initial depth is
$L_W/h_0 = 2 \cosh^{-1}(\sqrt{2})/\sqrt{\mu_0}$.
We seek to compare this bulk shape parameter defined at each point in
time $t$ with the local parameters defined at each point in space.
Therefore, we define $L_W(x) = L_W[\tpeak(x)]$ at the
time $\tpeak(x)$ when the wave peak passes location $x$.

\section{\label{sec:results}Results}
\noindent
Now, we use the results of the numerical simulations to investigate the
effect of wind on solitary wave shoaling and shape.
We will present shape statistics (\cref{sec:shape_params}) for the 20
different runs (\cref{tab:parameters}) to detail the wave-shape changes
and prebreaking behavior across the parameter space.
For the remainder of the paper we will return to dimensional variables.

\subsection{Profiles of shoaling solitary waves with wind}
\begin{figure}
  \centering
  { 
    \phantomsubfloat{\label{fig:snapshots_solitary:a}}
    \phantomsubfloat{\label{fig:snapshots_solitary:b}}
    \phantomsubfloat{\label{fig:snapshots_solitary:c}}
    \phantomsubfloat{\label{fig:snapshots_solitary:d}}
    \phantomsubfloat{\label{fig:snapshots_solitary:e}}
    \phantomsubfloat{\label{fig:snapshots_solitary:f}}
    \phantomsubfloat{\label{fig:snapshots_solitary:g}}
    \phantomsubfloat{\label{fig:snapshots_solitary:h}}
  }
  \includegraphics{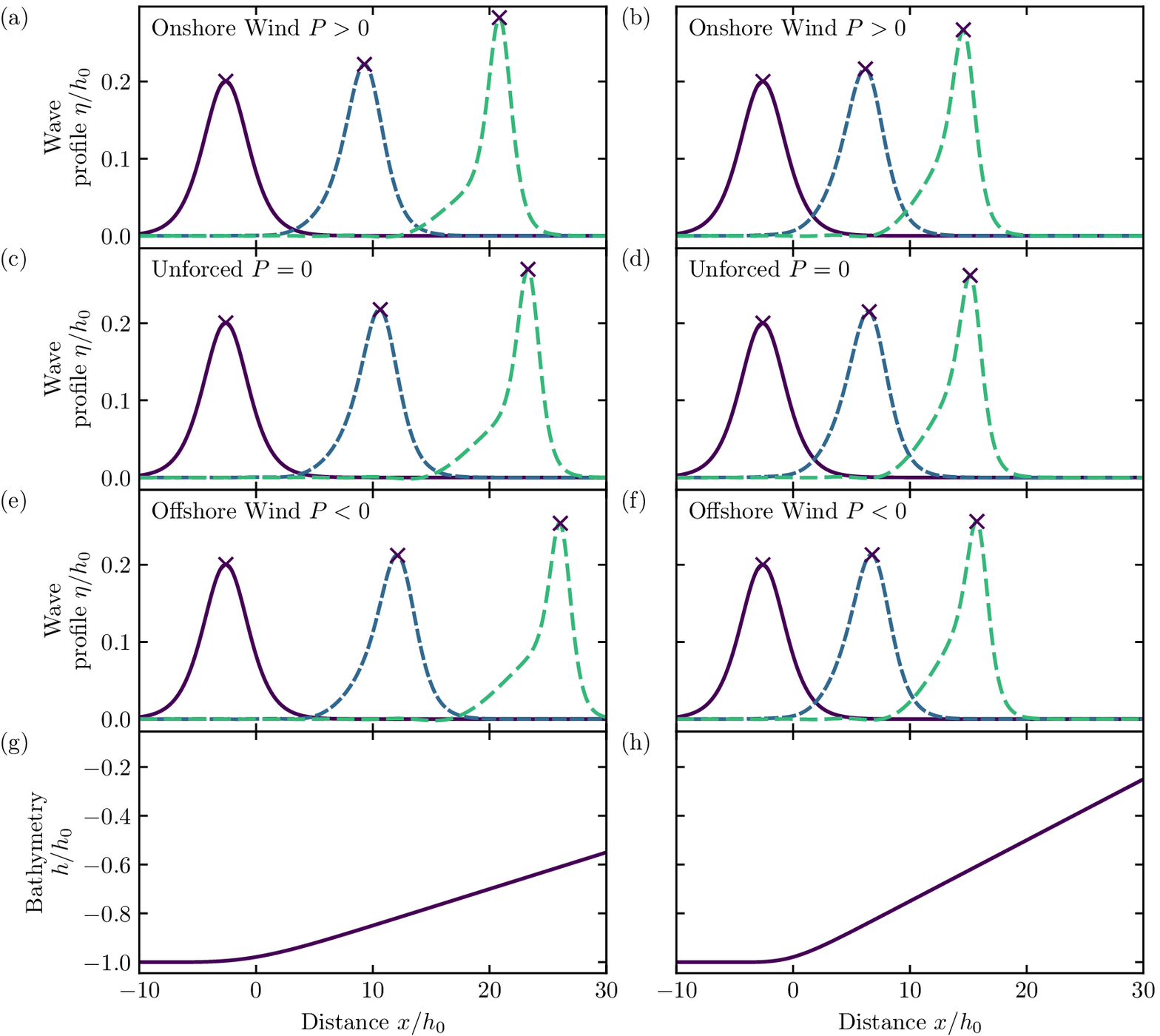}
  \caption{%
    Shoaling solitary-wave $\eta$ evolution under
    [\protect\subref{fig:snapshots_solitary:a},%
    \protect\subref{fig:snapshots_solitary:b}]
    onshore $P>0$,
    [\protect\subref{fig:snapshots_solitary:c},%
    \protect\subref{fig:snapshots_solitary:d}]
    unforced $P = 0$,
    and
    [\protect\subref{fig:snapshots_solitary:e},%
    \protect\subref{fig:snapshots_solitary:f}]
    offshore $P<0$ wind-induced surface pressure
    versus nondimensional distance $x/h_0$
    as the wave propagates up the
    [\protect\subref{fig:snapshots_solitary:g},%
    \protect\subref{fig:snapshots_solitary:h}]
    planar bathymetry.
    The profile times shown depend on the Froude number
    \labelcref{eq:froude_num} and therefore vary between the panels.
    The first profile (purple) occurs when the peak is located at
    $x=-L_0$ where the pressure begins turning on and the time is
    defined so $t=0$ here.
    The last profile (green) occurs when the convective prebreaking
    condition $\max_x(\Fr) = \Frpb = 1/3$ is met (\cf{}
    \cref{sec:breaking_criteria}), and the middle profile (blue) occurs
    at a time halfway between the first and last profiles.
    Both columns have $\epsilon_0=0.2$ and $\mu_0 = 0.15$, and
    [\protect\subref{fig:snapshots_solitary:a},%
    \protect\subref{fig:snapshots_solitary:e}]
    the left-column forced cases have $\abs{P /(\rho_w g L_0
    \epsilon_0)} = 0.05$ and $\beta = 0.015$ while
    [\protect\subref{fig:snapshots_solitary:b},%
    \protect\subref{fig:snapshots_solitary:f}]
    the right-column forced cases use $\abs{P /(\rho_w g L_0
    \epsilon_0)} = 0.025$ and $\beta = 0.025$.
    The $\times$'s denote the locations with the highest Froude number
    \labelcref{eq:froude_num}, and the $\times$'s on the last
    profiles (green) are the prebreaking locations $\xpb$.
    We only display a subset of the full spatial domain.
    Note that the aspect ratio is chosen to highlight the wind's effect
    on the shoaling solitons.
  }\label{fig:snapshots_solitary}
\end{figure}

First, we qualitatively investigate the effect of varying pressures $P$
and bathymetric slopes $\beta$ on solitary-wave shoaling by examining
the wave profile $\eta/h_0$, normalized by the initial depth $h_0$, at
three different times $t$ (\cref{fig:snapshots_solitary}) corresponding
to when the solitary wave first feels the slope ($t=0$), the time of
prebreaking ($t=\tpb$), and half-way between ($t=\tpb/2$).
Note that these $t=0$ wave profiles (purple in
\cref{fig:snapshots_solitary}) are nearly identical to the
$\sech^2(x/L_0)$ initial condition \labelcref{eq:initial_condition} since the
waves have only propagated over a flat bottom
[\cref{fig:snapshots_solitary:g,fig:snapshots_solitary:h}] and the
pressure has not yet been turned on.
Halfway to prebreaking ($t=\tpb/2$, blue), the solitary wave has grown
through shoaling with a steeper front face ($+x$ side) and increased
asymmetry for all $P$ and $\beta$.
At the time of prebreaking ($t=\tpb$, green) the solitary wave has
increased in height, steepened, and gained a substantial rear shelf
for all $P$ and $\beta$.
These changes are likely even larger for waves propagating to full wave
breaking ($\Fr \approx 1$).
The generation of rear shelves by shoaling solitary waves as in
\cref{fig:snapshots_solitary} was first calculated by
\citet{miles1979korteweg} and resulted from the mass shed by the
$\sech^2$ wave as it narrowed.
Onshore wind ($P>0$) reinforces the shoaling-based wave
growth and yields relatively narrow peak widths for both $\beta$
[\cref{fig:snapshots_solitary:a,fig:snapshots_solitary:b}].
In contrast, offshore wind ($P<0$) reduces the wave shoaling, but
results in wider peak widths
[\cref{fig:snapshots_solitary:e,fig:snapshots_solitary:f}].
For instance, the width at prebreaking on the milder slope
$\beta=0.015$ is $L_W/h(x) = \num{3.75}$ under onshore winds and
$L_W/h(x) = \num{4.26}$ under offshore winds.
These differences in wave-shoaling result in the offshore-forced ($P<0$)
solitary wave reaching prebreaking ($\xpb$, $\times$'s in
\cref{fig:snapshots_solitary}) farther onshore (shallower water) than
the onshore-forced ($P>0$) solitary wave.
For the $\beta = 0.015$ case, the onshore-forced wave reaches
prebreaking at $\xpb/h_0 = \num{20.8}$ while the offshore-forced wave
reaches prebreaking farther offshore at $\xpb/h_0 = \num{26.1}$.
Similarly, the larger beach slope [$\beta = 0.025$,
\cref{fig:snapshots_solitary:b,%
fig:snapshots_solitary:d,fig:snapshots_solitary:f}] causes waves to
reach $\xpb$ in less horizontal distance, though they prebreak in
shallow water than the milder beach slope ($\beta = 0.015$) waves.
For the unforced case, the $\beta = 0.015$ wave prebreaks at $\xpb/h_0
= \num{23.3}$ while the steeper beach $\beta = 0.025$ causes
prebreaking at $\xpb/h_0 = \num{15.2}$.
At $t=\tpb$, the rear shelf is wider and extends higher up the rear
face for offshore winds [$\approx 0.1 h_0$ in
\cref{fig:snapshots_solitary:e}] than for onshore winds
[$\approx 0.07 h_0$ in \cref{fig:snapshots_solitary:a}].
Again, we expect these differences to be even larger for fully breaking
waves ($\Fr \approx 1$).
As the control case, the unforced ($P=0$) solitary wave has $\xpb$
located between the onshore and offshore wind cases with an intermediate
rear shelf.
Finally, the milder slope ($\beta = 0.015$) has a sharper, more
pronounced rear shelf while the steeper slope ($\beta = 0.025$) has a
more gently sloping rear shelf.

\begin{figure}
  \centering
  { 
    \phantomsubfloat{\label{fig:snapshots_solitary_slope_velocity:a}}
    \phantomsubfloat{\label{fig:snapshots_solitary_slope_velocity:b}}
    \phantomsubfloat{\label{fig:snapshots_solitary_slope_velocity:c}}
    \phantomsubfloat{\label{fig:snapshots_solitary_slope_velocity:d}}
    \phantomsubfloat{\label{fig:snapshots_solitary_slope_velocity:e}}
    \phantomsubfloat{\label{fig:snapshots_solitary_slope_velocity:f}}
    \phantomsubfloat{\label{fig:snapshots_solitary_slope_velocity:g}}
    \phantomsubfloat{\label{fig:snapshots_solitary_slope_velocity:h}}
  }
  \includegraphics{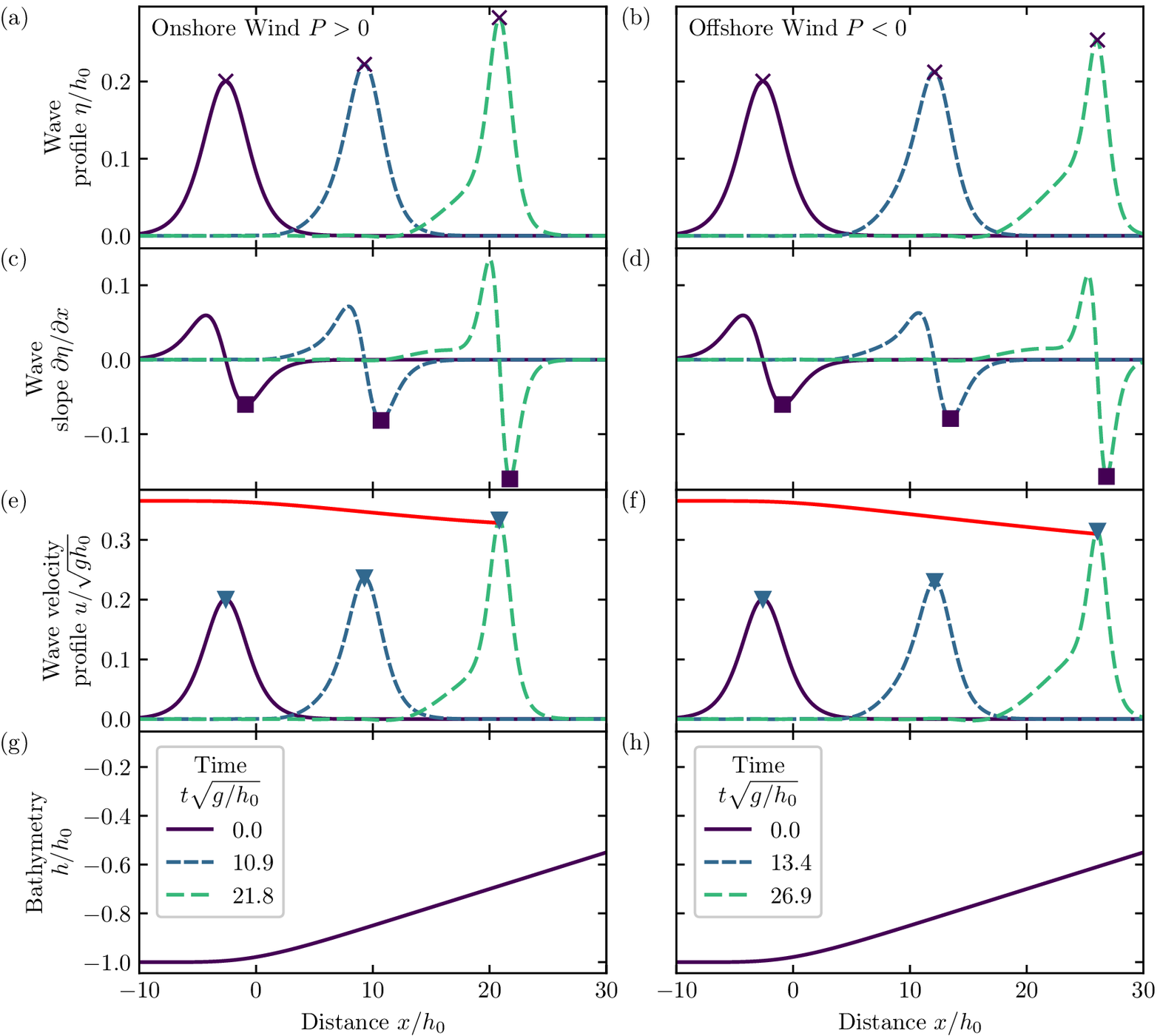}
  \caption{%
    Shoaling solitary-wave
    [\protect\subref{fig:snapshots_solitary_slope_velocity:a},%
    \protect\subref{fig:snapshots_solitary_slope_velocity:b}]
    nondimensional profile $\eta/h_0$,
    [\protect\subref{fig:snapshots_solitary_slope_velocity:c},%
    \protect\subref{fig:snapshots_solitary_slope_velocity:d}]
    slope $\partial \eta/\partial x$,
    and
    [\protect\subref{fig:snapshots_solitary_slope_velocity:e},%
    \protect\subref{fig:snapshots_solitary_slope_velocity:f}]
    nondimensional wave velocity profile $u/\sqrt{g h_0}$
    under
    [\protect\subref{fig:snapshots_solitary_slope_velocity:a},%
    \protect\subref{fig:snapshots_solitary_slope_velocity:c},%
    \protect\subref{fig:snapshots_solitary_slope_velocity:e}]
    onshore and
    [\protect\subref{fig:snapshots_solitary_slope_velocity:b},%
    \protect\subref{fig:snapshots_solitary_slope_velocity:d},%
    \protect\subref{fig:snapshots_solitary_slope_velocity:f}]
    offshore wind-induced surface pressure
    as the wave propagates up the
    [\protect\subref{fig:snapshots_solitary_slope_velocity:g},%
    \protect\subref{fig:snapshots_solitary_slope_velocity:h}]
    planar bathymetry.
    All panels use the same
    [\protect\subref{fig:snapshots_solitary_slope_velocity:g},%
    \protect\subref{fig:snapshots_solitary_slope_velocity:h}]
    bathymetry and differ only in the sign of the pressure forcing.
    Values are shown versus nondimensional distance $x/h_0$ for
    $\epsilon_0=0.2$, $\mu_0 = 0.15$, $\abs{P /(\rho_w g L_0 \epsilon_0)} =
    0.05$, $\beta=0.015$, and nondimensional times $t \sqrt{gh}/L_0$
    indicated in the legends.
    The red lines in
    [\protect\subref{fig:snapshots_solitary_slope_velocity:e},%
    \protect\subref{fig:snapshots_solitary_slope_velocity:f}]
    represent the phase speed $\cadi$ \labelcref{eq:c_adiabatic} at
    each location multiplied by the prebreaking Froude number
    $\Frpb = 1/3$.
    The $\times$'s denote the locations with the highest Froude number and
    the $\times$'s on the last (green) profiles are the prebreaking
    locations $\xpb$.
    The squares are the locations of the maximum slope magnitude
    $\abs{\partial \eta/\partial x}$ and the upside-down triangles
    represent the locations of the maximum wave velocity profile.
    We only display a subset of the full spatial domain.
    Note that the aspect ratio is chosen to highlight the wind's effect
    on the shoaling solitons.
  }\label{fig:snapshots_solitary_slope_velocity}
\end{figure}

We next investigate the impact of
onshore
[\cref{fig:snapshots_solitary_slope_velocity:a,%
fig:snapshots_solitary_slope_velocity:c,%
fig:snapshots_solitary_slope_velocity:e}]
and offshore
[\cref{fig:snapshots_solitary_slope_velocity:b,%
fig:snapshots_solitary_slope_velocity:d,%
fig:snapshots_solitary_slope_velocity:f}]
wind on shoaling waves' slopes $\partial_x \eta$ and wave velocity
profiles $u/\sqrt{g h_0}$.
\Cref{fig:snapshots_solitary_slope_velocity} highlights the influence of
wind directionality on wave shape by keeping the bathymetry and pressure
magnitude fixed while changing the pressure forcing's sign.
The wave slope
[\cref{fig:snapshots_solitary_slope_velocity:c,%
fig:snapshots_solitary_slope_velocity:d}]
highlights the shoaling- and wind-induced shape changes by accentuating
the front-rear asymmetry.
At $t=0$ [purple in
\cref{fig:snapshots_solitary_slope_velocity:a,%
fig:snapshots_solitary_slope_velocity:b}],
the wave slope has odd-parity about the peak.
However, as the wave propagates onshore, both the front and rear faces
steepen, though the front face steepens more dramatically.
The influence of the wind is most noticeable in three aspects: the
offshore-forced wave [$P=-0.05$,
\cref{fig:snapshots_solitary_slope_velocity:b}] is \SI{10}{\percent}
smaller than the onshore forced wave [$P=0.05$,
\cref{fig:snapshots_solitary_slope_velocity:a}]; the
offshore-forced rear-face wave slope
[\cref{fig:snapshots_solitary_slope_velocity:d}] is \SI{16}{\percent}
smaller than the onshore-forced wave slope
[\cref{fig:snapshots_solitary_slope_velocity:c}], though the front-face
slope is only \SI{2}{\percent} smaller;
and the trailing shelf's slope extends further behind the
offshore-forced wave [$\approx 8 h_0$,
\cref{fig:snapshots_solitary_slope_velocity:d}] than the onshore-forced
wave [$\approx 5 h_0$, \cref{fig:snapshots_solitary_slope_velocity:c}].
The wave velocity profile $u/\sqrt{g h_0}$
[\cref{eq:water_velocity},
\cref{fig:snapshots_solitary_slope_velocity:e,%
fig:snapshots_solitary_slope_velocity:f}]
nearly mirrors the wave profile
[\cref{fig:snapshots_solitary_slope_velocity:a,%
fig:snapshots_solitary_slope_velocity:b}],
as is expected given that $u \propto \eta$ to leading order
\labelcref{eq:water_velocity}.
Finally, the phase speed $\cadi$ [red, \cref{eq:c_adiabatic}]
decreases as the wave shoals which enhances convective prebreaking,
though $\cadi$ only varies \SI{3}{\percent} between the onshore and
offshore winds.
Note, in
\cref{fig:snapshots_solitary_slope_velocity:e,%
fig:snapshots_solitary_slope_velocity:f},
$\cadi$ is multiplied by $\Frpb=1/3$ so that the intersection of the red
curve with the wave velocity profile occurs at $\xpb$, the location of
prebreaking.
We highlight that the prebreaking quantity $u/c \approx \Fr = 1/3$ is
smaller than the unified breaking onset criteria $u/c = 0.85$ described
by \citet{derakhti2020unified}.

\subsection{\label{sec:pb_width_results}Shape statistics with shoaling and variations of prebreaking zone
width with wind}
\begin{figure}
  \centering
  { 
    \phantomsubfloat{\label{fig:statistics_solitary:a}}
    \phantomsubfloat{\label{fig:statistics_solitary:b}}
    \phantomsubfloat{\label{fig:statistics_solitary:c}}
    \phantomsubfloat{\label{fig:statistics_solitary:d}}
    \phantomsubfloat{\label{fig:statistics_solitary:e}}
    \phantomsubfloat{\label{fig:statistics_solitary:f}}
    \phantomsubfloat{\label{fig:statistics_solitary:g}}
  }
  \includegraphics{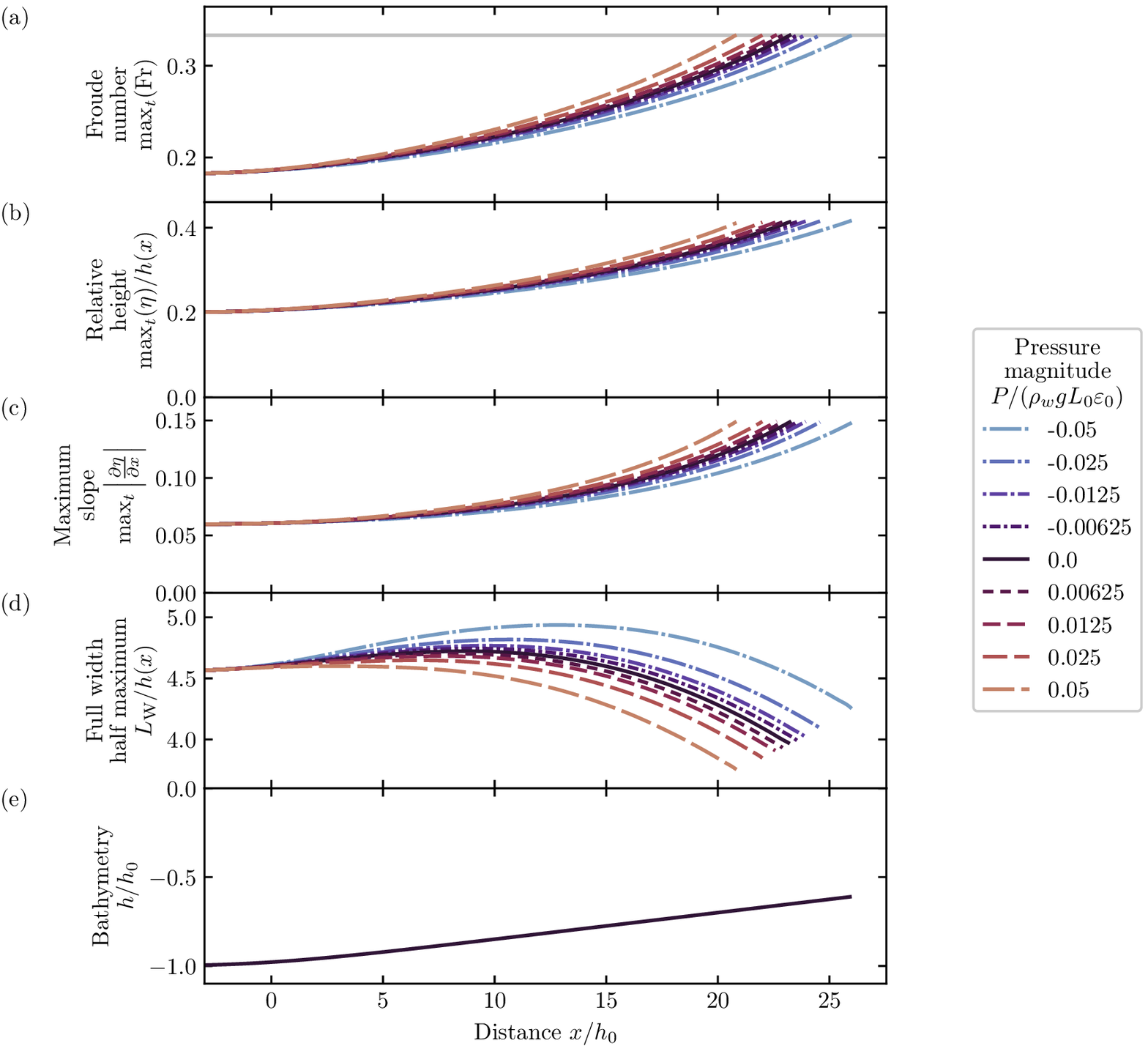}
  \caption{%
    Shoaling solitary-wave shape statistics under onshore and offshore
    pressure forcing versus nondimensional distance $x/h_0$.
    The
    \protect\subref{fig:statistics_solitary:a}
    Froude number $\max_t(\Fr)$ \labelcref{eq:froude_num},
    \protect\subref{fig:statistics_solitary:b}
    maximum height normalized by the local water depth
    $\max_t(\eta)/h(x)$,
    \protect\subref{fig:statistics_solitary:c}
    maximum slope $\max_t(\abs{\partial \eta/\partial x})$,
    and
    \protect\subref{fig:statistics_solitary:d}
    full width at half maximum normalized by the local water depth
    $L_W/h(x)$ (\cf{} \cref{sec:shape_params})
    are displayed at each location along the
    \protect\subref{fig:statistics_solitary:g}
    planar bathymetry.
    Results are shown for $\epsilon_0=0.2$, $\mu_0 = 0.15$,
    $\beta=0.015$, and pressure magnitude $\abs{P /(\rho_w g L_0
    \epsilon_0)}$ up to $0.05$, as indicated in the legend.
    The solid black line is the unforced case, $P = 0$.
    The light gray line in \protect\subref{fig:statistics_solitary:a}
    represents the convective prebreaking Froude number
    $\Frpb = 1/3$ at which the simulations were stopped.
  }\label{fig:statistics_solitary}
\end{figure}

Building on the previous qualitative descriptions of the wave
shape, slope, and velocity, we now quantify the change
in the shoaling wave's shape parameters for onshore and offshore
$P$~(\cref{fig:statistics_solitary}).
First, we consider the maximum Froude number $\max_t(\Fr)$ as a function
of nondimensional position $x/h_0$ [\cref{fig:statistics_solitary:a}].
In the flat region ($x<0$), the maximum Froude number is $\max_t(\Fr) =
\num{0.1986}$, and it increases as the waves shoal to the prebreaking
value $\max_t(\Fr) = \Frpb = 1/3$ (light gray line).
The wind has a significant impact on the location of prebreaking
$\xpb$, with onshore wind (red) causing the Froude number to increase
faster and $\xpb$ to occur farther offshore than offshore wind (blue)
does.
This can also be seen in
\cref{fig:snapshots_solitary_slope_velocity:e,%
fig:snapshots_solitary_slope_velocity:f},
where the maximum velocities $u/\sqrt{g h_0}$ (upside-down triangles),
which are proportional to $\max_t(\Fr)$, are growing faster for the
onshore wind [\cref{fig:snapshots_solitary_slope_velocity:e}] than
the offshore wind [\cref{fig:snapshots_solitary_slope_velocity:f}].
Notably, at a fixed location $x/h_0$, the $\max_x(\Fr)$ varies
substantially (\eg \numrange[range-phrase = \text{~to~}]{0.28}{0.32} at
$x/h_0 = 20$).
In addition, we consider the maximum height $\max_t(\eta)$ at a fixed
location and normalized by the local water depth $h(x)$
[\cref{fig:statistics_solitary:b}].
For all pressures $P$, the solitary wave increases in height, but the
onshore wind enhances this growth while the offshore wind partially
suppresses the growth.
Again, this is apparent in the evolution of the maximums
$\eta(\xpeak)/h_0$ in \cref{fig:snapshots_solitary}, with the peak
locations $\xpeak$ closely approximated by the $\times$'s marking the
location of maximum $\Fr$.
Since $\Fr \propto \eta$ to leading order, the relative height
at prebreaking is approximately \num{0.41} for all $P$
[\cref{fig:statistics_solitary:b}] with offshore-forced waves
slightly larger (\SI{1}{\percent}) than onshore-forced waves.

\Cref{fig:statistics_solitary:c} shows the evolution of the maximum wave
slope magnitude $\max_t \abs{\partial_x \eta}$, corresponding to the
front face's slope [\cref{fig:snapshots_solitary_slope_velocity:c,%
fig:snapshots_solitary_slope_velocity:d}].
Like the relative height [\cref{fig:statistics_solitary:b}], the
steepness is enhanced by onshore wind ($P>0$), suppressed for offshore
wind ($P<0$), and approaches nearly the same prebreaking value of
\num{0.15} for all wind speeds, being only \SI{1}{\percent} larger for
onshore winds than offshore winds.
The maximum slope at prebreaking is nearly constant because solitary
waves have a fixed relationship between the wave height and wave
width (and hence slope) as discussed in \cref{sec:ICs}.
And the height at prebreaking is approximately constant since $\Fr
\propto \eta$ to leading order and $\Fr = \Frpb$ is constant.
However, this relationship is only approximate on a slope, with
deviations due to nonlinearity, dispersion, shoaling, and wind
forcing~\labelcref{eq:froude_terms}.
Finally, we examine the FWHM $L_W$, normalized by the local water depth
$h(x)$ [\cref{fig:statistics_solitary:d}].
While $L_W/h(x)$ ultimately decreases from its initial value of
\num{4.56} for all pressure magnitudes, there is significant variation
in the prebreaking value.
For our parameters, $L_W/h(x)$ changes \SI{215}{\percent} more for
onshore wind ($P=0.05$) than offshore wind ($P=-0.05$) from start to
prebreaking.
\Cref{fig:snapshots_solitary_slope_velocity:a,fig:snapshots_solitary_slope_velocity:b}
show that the rear shelf does not rise to half the wave height, so
the FWHM does not incorporate the shelf's width.
Instead, the onshore-forced narrowing is occurring in the top region
above the shelf.
Hence, while the relative height and slope at prebreaking are largely
similar for all the wind speeds, the FWHM at prebreaking is strongly
affected by the wind speed indicating wind effects on shoaling shape.
We expect the wind-induced changes to maximum wave slope and FWHM to be
even more stronger approaching wave breaking ($\Fr \approx 1$).

\begin{figure}
  \centering
  \includegraphics{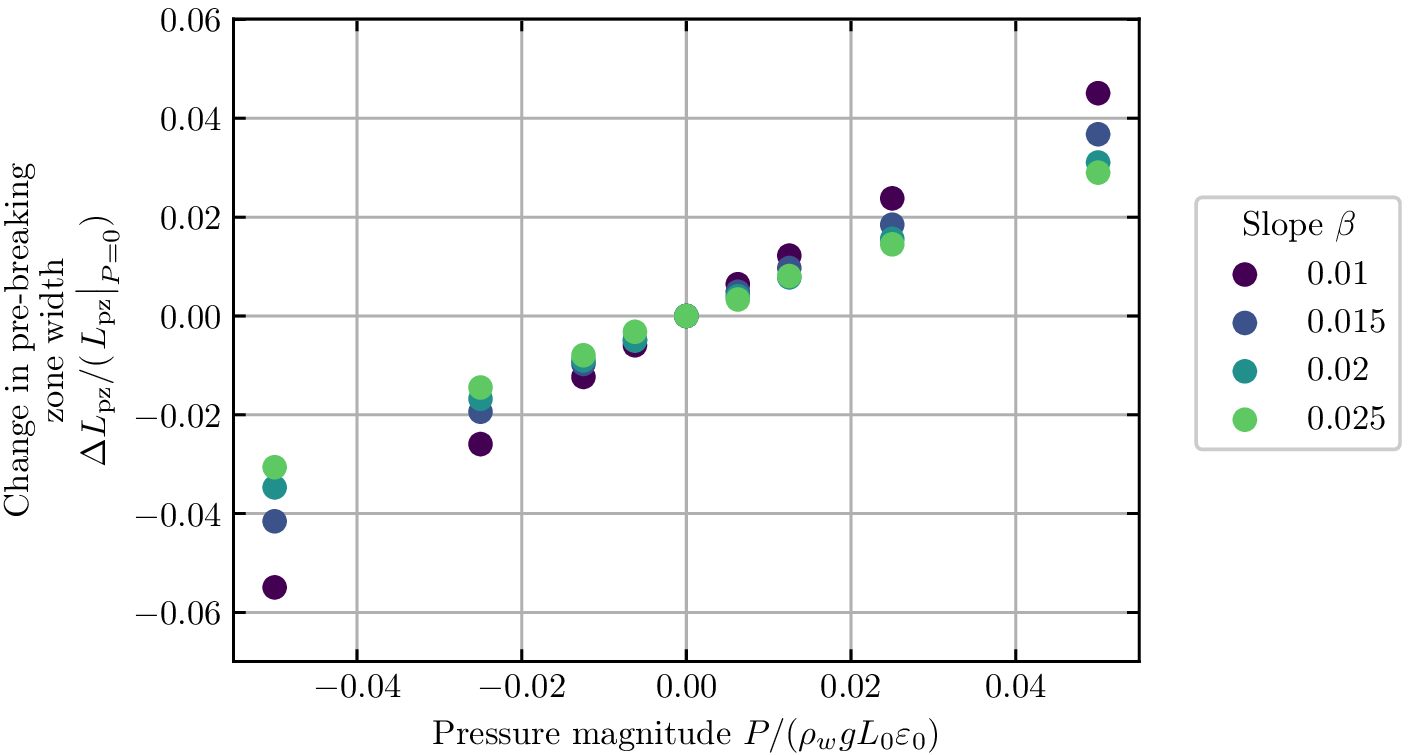}
  \caption{%
    The fractional change in prebreaking zone width $\Delta \Lpz$
    compared to the unforced case $\eval{\Lpz}_{P=0}$ (\cf{}
    \cref{sec:shape_params}) versus the nondimensional pressure
    magnitude $P/(\rho_w g L_0 \epsilon_0)$.
    The results are shown for beach slopes $\beta =
    \numrange[range-phrase={\text{--}}]{0.01}{0.025}$ as indicated in
    the legend.
  }\label{fig:break_locations}
\end{figure}

We also investigate the change in the prebreaking zone width $\Delta
\Lpz$ (\cref{sec:shape_params}) as a function of pressure $P/(\rho_w g
L_0 \epsilon_0)$ for four different values of the beach slope
$\beta$~(\cref{fig:break_locations}).
First, $\Delta \Lpz$ is linearly related to the pressure magnitude, and
the wind has a larger effect on $\Delta \Lpz$ for smaller beach slopes,
with $P/(\rho_w g L_0 \epsilon_0) = -0.05$ changing the prebreaking
zone width by approximately \SI{5}{\percent} for the smallest slope
$\beta=0.01$.
This is because the wind has more time to affect the wave before it
reaches prebreaking.
This wind-induced change in prebreaking location is visible in
\cref{fig:snapshots_solitary}, where the prebreaking location $\xpb$
($\times$'s on green profiles) occurs closer to the shoreline ($+x$
direction) for offshore winds $P<0$ [\cref{fig:snapshots_solitary:e,%
fig:snapshots_solitary:f}] than for onshore winds $P>0$
[\cref{fig:snapshots_solitary:a,fig:snapshots_solitary:b}].
Additionally, we note that for the smallest slope $\beta=0.01$, the
fractional change in prebreaking zone width $\Delta
\Lpz/(\eval{\Lpz}_{P=0})$ is asymmetric with respect to pressure, with
offshore $P/(\rho_w g L_0 \epsilon_0) = -0.05$ yielding a
\SI{22}{\percent} larger change than onshore $P/(\rho_w g L_0
\epsilon_0) = 0.05$~(\cref{fig:break_locations}).
For fully breaking waves, the wind-induced changes to the breaking-zone
width $\Delta \Lbz$ would likely be larger and the unforced surf zone
width $\eval{\Lbz}_{P=0}$ would be smaller (as waves propagate closer to
shore before fully breaking), making the fractional change $\Delta
\Lbz/(\eval{\Lbz}_{P=0})$ much larger.

\subsection{Normalized prebreaking wave shape changes induced by wind
and shoaling}
\begin{figure}
  \centering
  { 
    \phantomsubfloat{\label{fig:wave_shapes:a}}
    \phantomsubfloat{\label{fig:wave_shapes:b}}
    \phantomsubfloat{\label{fig:wave_shapes:c}}
  }
  \includegraphics{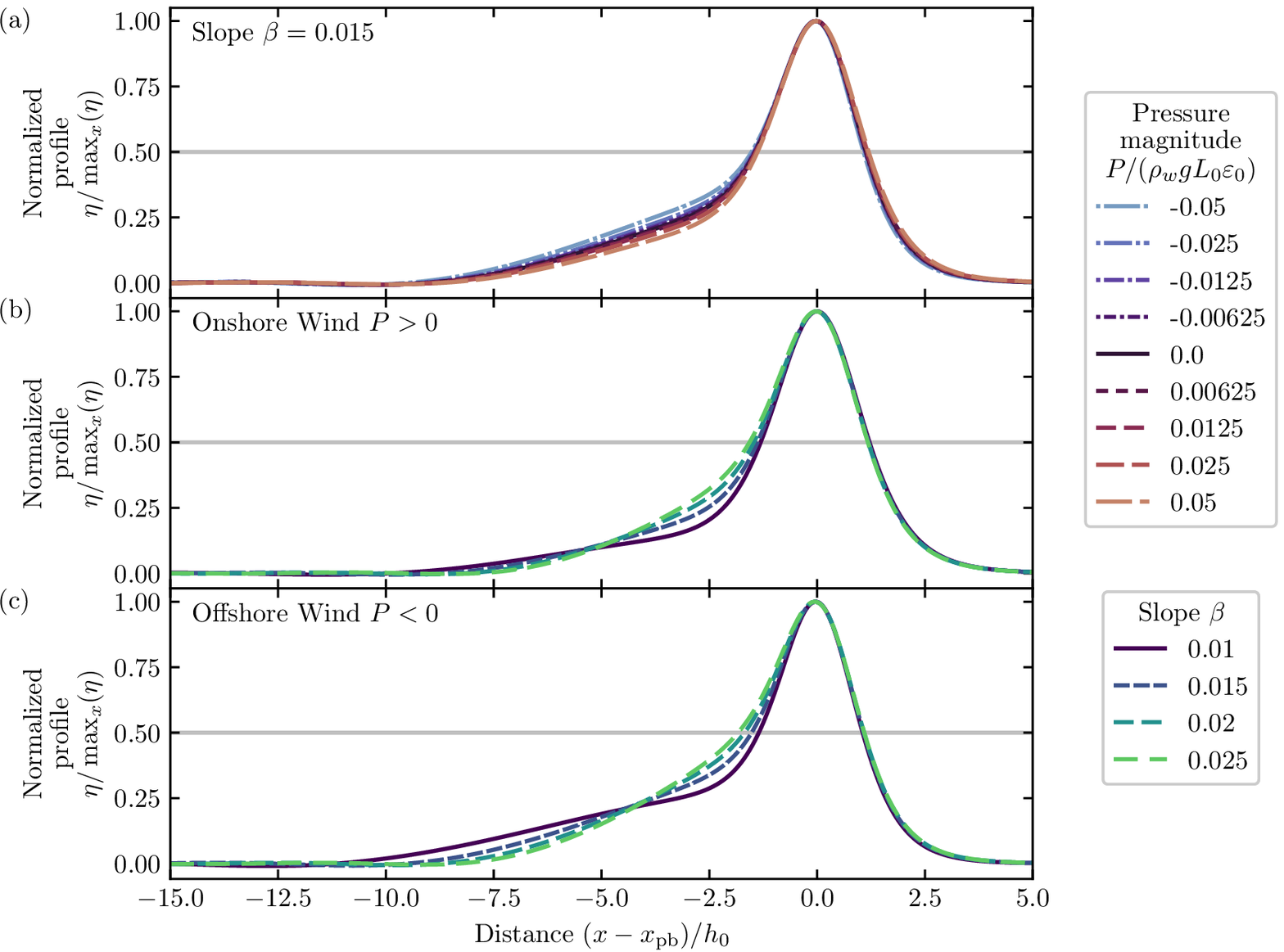}
  \caption{%
    Prebreaking wave profile $\eta/\max_x(\eta)$
    normalized by the maximum height versus nondimensional position
    $(x-\xpb)/h_0$ relative to the prebreaking location $\xpb$.
    All profiles occur at prebreaking $\tpb$ when $\max_x(\Fr) = \Frpb
    = 1/3$ (\cf{} \cref{sec:breaking_criteria}) and display different
    values of the \protect\subref{fig:wave_shapes:a}
    pressure magnitude $P / (\rho_w g L_0 \epsilon_0)$
    and the
    \protect\subref{fig:wave_shapes:b},%
    \protect\subref{fig:wave_shapes:c}
    bottom slope $\beta$, as indicated in the legend.
    Results are shown for $\epsilon_0=0.2$, $\mu_0 = 0.15$, and
    \protect\subref{fig:wave_shapes:a}
    slope $\beta = 0.015$,
    \protect\subref{fig:wave_shapes:b}
    onshore $P / (\rho_w g L_0 \epsilon_0) = 0.05$,
    or
    \protect\subref{fig:wave_shapes:c}
    offshore $P / (\rho_w g L_0 \epsilon_0) = -0.05$
    pressure magnitude.
    The light gray line shows where the FWHM is measured.
  }\label{fig:wave_shapes}
\end{figure}

As \cref{fig:statistics_solitary} quantified the shape statistics at
prebreaking for all $x$, we now directly investigate the effect of
pressure $P$ and shoaling $\beta$ on prebreaking wave shape by
normalizing each prebreaking wave profile $\eta$ by its maximum height
$\max_x(\eta)$ and aligning the prebreaking locations
$\xpb/h_0$~(\cref{fig:wave_shapes}).
Each solution is dominated by the $\sech^2$ wave centered near
$x-\xpb=0$, which becomes taller and narrower as the wave shoals as
required by energy conservation~\citep{miles1979korteweg}.
Furthermore, while the $\sech^2$ component is symmetric in time at a
fixed location, it becomes slightly deformed when viewed at a fixed time
as the front face moves slower than the rear
face~\citep{newell1985solitons,knickerbocker1985propagation}.
We also observe a shelf behind the wave, which
\citet{miles1979korteweg} calculated by requiring that the right-moving
mass-flux be conserved as the $\sech^2$ narrows and sheds mass.
While long-duration calculations of the Miles shelf reveal a nearly
horizontal shelf extending far behind the
wave~\citep{knickerbocker1980shelves,knickerbocker1985propagation}, our
shelf instead slopes gently downward, likely due to insufficient
development time and distance.

In \cref{fig:wave_shapes}, we plot the prebreaking wave shape for fixed
bottom slope $\beta$ [\cref{fig:wave_shapes:a}] and fixed pressure
magnitude $P$ [\cref{fig:wave_shapes:b,fig:wave_shapes:c}].
For a fixed slope [\cref{fig:wave_shapes:a}], the front wave
faces at prebreaking are qualitatively very similar and match an
unforced solitary wave of the same height.
However, wind strongly affects the rear shelves as observed in
\cref{fig:snapshots_solitary}.
The offshore winds (blue) cause the shelf to be thicker and extend
higher up the rear wave face than the offshore wind (reds) do, although
the shelf intersects $z=0$ at $(x-\xpb)/h_0 \approx -10$ for all wind
speeds.
Nevertheless, all of the changes in \cref{fig:wave_shapes} would likely
be enhanced for fully breaking waves as wind and shoaling effects have
longer to act on the wave.

We also consider the wave shape at breaking for different values of the
beach slope $\beta$ with a fixed onshore
[\cref{fig:wave_shapes:b}] or offshore
[\cref{fig:wave_shapes:c}] wind.
The rear half of the wave shows that bottom slope $\beta$ affects the
rear shelf differently than pressure $P/(\rho_w g L_0 \epsilon_0)$ does.
While the shelf intersected $z=0$ at the same location for
all wind speeds [\cref{fig:wave_shapes:a}], increasing $\beta$ causes
the intersection point (\ie the base of the shelf) to move forward and
closer to the peak.
Finally, the offshore wind [\cref{fig:wave_shapes:c}] causes a
noticeably larger shelf than the onshore wind
[\cref{fig:wave_shapes:b}] for the weakest slope $\beta = 0.01$
(purple), with a similar pattern observed in
\cref{fig:snapshots_solitary_slope_velocity:a} ($\beta=0.015$) compared
to \cref{fig:snapshots_solitary_slope_velocity:b} ($\beta=0.025$).
However, this difference is much smaller for the steeper (green) slopes,
implying that stronger shoaling partially suppresses the wind-induced
shape change because there is less time for pressure to act prior to
prebreaking.

\section{\label{sec:discussion}Discussion}
\subsection{\label{sec:press_mag}Wind speed}
\noindent
Our derivation in \cref{sec:derivation} coupled wind to the wave's
motion through the use of a surface pressure \cref{eq:press_def}.
The resulting vKdV--Burgers equation \labelcref{eq:vckdvb_lab_frame} has a
wind-induced term dependent on the pressure magnitude constant
$P/(\rho_w g L_0 \epsilon_0)$.
We analyze the evolution and prebreaking of solitary waves for
variable $P$ (\cref{sec:results}).
While the usage of $P$ was the most natural since it is the physical
coupling between wind and waves (in the absence of viscous tangential
stress), measuring the surface pressure is challenging in field
observations or laboratory
experiments~\citep{donelan2006wave,buckley2019turbulent}.
Therefore, we also consider the evolution and prebreaking of the
shoaling solitary waves as a function of the wind speed $U$.
\Citet{zdyrski2021wind} considered a surface pressure acting on a
flat-bottom KdV solitary wave initial condition [equivalent to our
\cref{eq:initial_condition}] with dimensional form
\begin{equation}
  \eta = \epsilon h \sech^2\pqty{\sqrt{\frac{3 \epsilon}{4}}
    \frac{x}{h}}^2 \,,
  \label{eq:initial_condition_dim}
\end{equation}
with nondimensional height $\epsilon = H/h$ and width $L = 2 h/\sqrt{3
\epsilon}$ in water of depth $h$.
\Citet{zdyrski2021wind} then related wind speed $U$ to the surface
pressure magnitude using
\begin{equation}
  \frac{U}{\sqrt{g h}} = 1 \pm
    \sqrt{\frac{1}{5} \abs{\frac{P}{\rho_w g h \epsilon}}
    \frac{\rho_w}{\rho_a} \frac{2}{4.91}}
  = 1 \pm
    \sqrt{\frac{1}{5} \abs{\frac{P}{\rho_w g L \epsilon}}
    \frac{\rho_w}{\rho_a} \frac{4}{4.91 \sqrt{3 \epsilon}}} \,,
  \label{eq:wind_speed}
\end{equation}
where $U$ is measured at a height of half the solitary wave's width.
This parametrization originated from observations by
\citet{donelan2006wave} of nonseparated wind forcing for periodic,
shallow-water waves and was adapted to solitary waves in shallow water
by \citet{zdyrski2021wind}.
The relationship is accurate for the wind and wave conditions of
\citet{donelan2006wave}, but should be interpreted qualitatively here.
Note that the radicand differs by a factor of \num{2} from
\citet{zdyrski2021wind} owing to the different definitions of $\epsilon$.
Even though \cref{eq:wind_speed} was originally applied to flat-bottomed
KdV solitary waves \labelcref{eq:initial_condition}, our assumption that
$\gamma = L/L_b \ll 1$ implies that the bathymetry is approximately flat
over the wave's width $2 L$.
Therefore, we use \cref{eq:wind_speed} to translate between the pressure
$P/(\rho_w g L_0 \epsilon_0)$ and the wind speed $U$ at any point on the
sloping bathymetry by using the local $\epsilon$ and $h$ and relating
the initial pressure to the local pressure $P/(\rho_w g L \epsilon) =
(\epsilon_0 L_0/\epsilon L) P/(\rho_w g L_0 \epsilon_0)$.

\begin{table}
  \centering
  \sisetup{round-mode=figures, round-precision=2, table-format=2.4}
  \newcommand\gVal{9.8}
  \newcommand\epsVal{0.2}
  \newcommand\rhoaOnrhow{0.0011225}
  \newcommand\HVal{1}
  \FPeval{\muOnEps}{3/4*\HVal}
  \caption{%
    Wind speeds as functions of pressure $P /(\rho_w g L \epsilon)$
    and local depth $h$ for solitary waves
    \labelcref{eq:initial_condition_dim} with $\epsilon = \epsVal$.
    $U_{\text{onshore}}$ corresponds to $P>0$ and $U_{\text{offshore}}$
    to $P<0$.
    The conversion from $P/(\rho_w g L \epsilon)$ to $U$ is given in
    \cref{eq:wind_speed}.
  }\label{tab:wind_speeds}
  \begin{spreadtab}{{tabular*}{\linewidth}{ S[table-number-alignment=left]@{\extracolsep{\fill}}
    S@{\extracolsep{\fill}} S@{\extracolsep{\fill}}
    S@{\extracolsep{\fill}} S[round-precision=1,table-format=1.0]@{\extracolsep{\fill}}
    S@{\extracolsep{\fill}} S@{\extracolsep{\fill}} }}
    \toprule \toprule
    {@ $\abs{P / (\rho_w g L \epsilon)}$}
               & {@ $h [\si{\meter}]$}
               & {@ $U_{\text{onshore}} [\si{\meter\per\second}]$}
               & {@ $U_{\text{offshore}} [\si{\meter\per\second}]$}
               & {@ $h [\si{\meter}]$}
               & {@ $U_{\text{onshore}} [\si{\meter\per\second}]$}
               & {@ $U_{\text{offshore}} [\si{\meter\per\second}]$}
               \\
    \midrule
    0
        & 2.5
        & (\gVal*[-1,0])^(0.5)*(1+(1/5*[-2,0]/\rhoaOnrhow*2/4.91/(\muOnEps*\epsVal
)^(0.5))^(0.5))
        & (\gVal*[-2,0])^(0.5)*(1-(1/5*[-3,0]/\rhoaOnrhow*2/4.91/(\muOnEps*\epsVal
)^(0.5))^(0.5))
        & 1
        & (\gVal*[-1,0])^(0.5)*(1+(1/5*[-5,0]/\rhoaOnrhow*2/4.91/(\muOnEps*\epsVal
)^(0.5))^(0.5))
        & (\gVal*[-2,0])^(0.5)*(1-(1/5*[-6,0]/\rhoaOnrhow*2/4.91/(\muOnEps*\epsVal
)^(0.5))^(0.5)) \\
    0.00625
        & 2.5
        & (\gVal*[-1,0])^(0.5)*(1+(1/5*[-2,0]/\rhoaOnrhow*2/4.91/(\muOnEps*\epsVal
)^(0.5))^(0.5))
        & (\gVal*[-2,0])^(0.5)*(1-(1/5*[-3,0]/\rhoaOnrhow*2/4.91/(\muOnEps*\epsVal
)^(0.5))^(0.5))
        & 1
        & (\gVal*[-1,0])^(0.5)*(1+(1/5*[-5,0]/\rhoaOnrhow*2/4.91/(\muOnEps*\epsVal
)^(0.5))^(0.5))
        & (\gVal*[-2,0])^(0.5)*(1-(1/5*[-6,0]/\rhoaOnrhow*2/4.91/(\muOnEps*\epsVal
)^(0.5))^(0.5)) \\
    0.0125
        & 2.5
        & (\gVal*[-1,0])^(0.5)*(1+(1/5*[-2,0]/\rhoaOnrhow*2/4.91/(\muOnEps*\epsVal
)^(0.5))^(0.5))
        & (\gVal*[-2,0])^(0.5)*(1-(1/5*[-3,0]/\rhoaOnrhow*2/4.91/(\muOnEps*\epsVal
)^(0.5))^(0.5))
        & 1
        & (\gVal*[-1,0])^(0.5)*(1+(1/5*[-5,0]/\rhoaOnrhow*2/4.91/(\muOnEps*\epsVal
)^(0.5))^(0.5))
        & (\gVal*[-2,0])^(0.5)*(1-(1/5*[-6,0]/\rhoaOnrhow*2/4.91/(\muOnEps*\epsVal
)^(0.5))^(0.5)) \\
    0.025
        & 2.5
        & (\gVal*[-1,0])^(0.5)*(1+(1/5*[-2,0]/\rhoaOnrhow*2/4.91/(\muOnEps*\epsVal
)^(0.5))^(0.5))
        & (\gVal*[-2,0])^(0.5)*(1-(1/5*[-3,0]/\rhoaOnrhow*2/4.91/(\muOnEps*\epsVal
)^(0.5))^(0.5))
        & 1
        & (\gVal*[-1,0])^(0.5)*(1+(1/5*[-5,0]/\rhoaOnrhow*2/4.91/(\muOnEps*\epsVal
)^(0.5))^(0.5))
        & (\gVal*[-2,0])^(0.5)*(1-(1/5*[-6,0]/\rhoaOnrhow*2/4.91/(\muOnEps*\epsVal
)^(0.5))^(0.5)) \\
    0.05
        & 2.5
        & (\gVal*[-1,0])^(0.5)*(1+(1/5*[-2,0]/\rhoaOnrhow*2/4.91/(\muOnEps*\epsVal
)^(0.5))^(0.5))
        & (\gVal*[-2,0])^(0.5)*(1-(1/5*[-3,0]/\rhoaOnrhow*2/4.91/(\muOnEps*\epsVal
)^(0.5))^(0.5))
        & 1
        & (\gVal*[-1,0])^(0.5)*(1+(1/5*[-5,0]/\rhoaOnrhow*2/4.91/(\muOnEps*\epsVal
)^(0.5))^(0.5))
        & (\gVal*[-2,0])^(0.5)*(1-(1/5*[-6,0]/\rhoaOnrhow*2/4.91/(\muOnEps*\epsVal
)^(0.5))^(0.5)) \\
    \bottomrule \bottomrule
  \end{spreadtab}
  \undef\gVal
  \undef\epsVal
  \undef\rhoaOnrhow
  \undef\HVal
\end{table}
\Cref{tab:wind_speeds} shows the onshore ($P>0$) and offshore ($P<0$)
wind speeds corresponding to the pressures used in our simulations for
two representative depths $h$.
It shows that the pressure magnitudes in our simulations correspond to
physically reasonable wind speeds, with onshore $U$ from
\SIrange[range-phrase = \text{~to~}]{3.1}{13}{\meter\per\second} for
water \SI{1}{\meter} deep or \SIrange[range-phrase =
\text{~to~}]{4.9}{20}{\meter\per\second} for water \SI{2.5}{\meter}
deep.
Notice that unforced waves with $P=0$ correspond to a wind speed
matching the wave phase speed $U = c$, with $c$ approximately
the linear shallow-water phase speed $c \approx \sqrt{g h}$.
In particular, this means that onshore $P>0$ and offshore $P<0$ winds
with the same pressure magnitude $\abs{P}$ will have different
wind-speed magnitudes $\abs{U}$.
Additionally, note that keeping $P$ fixed implies that the wind speed
$U$ changes as the wave shoals.
This is mostly due to the decrease in the phase speed $c \propto
\sqrt{gh}$, with higher-order effects coming from the $\epsilon$ and $L$
dependence of the radicand in \cref{eq:wind_speed}.
Finally, note that as the wave shoals and $\epsilon$ increases, the
height at which the wind speed is measured $z = L/2 = h/\sqrt{3
\epsilon}$ decreases.

\begin{figure}
  \centering
  \includegraphics{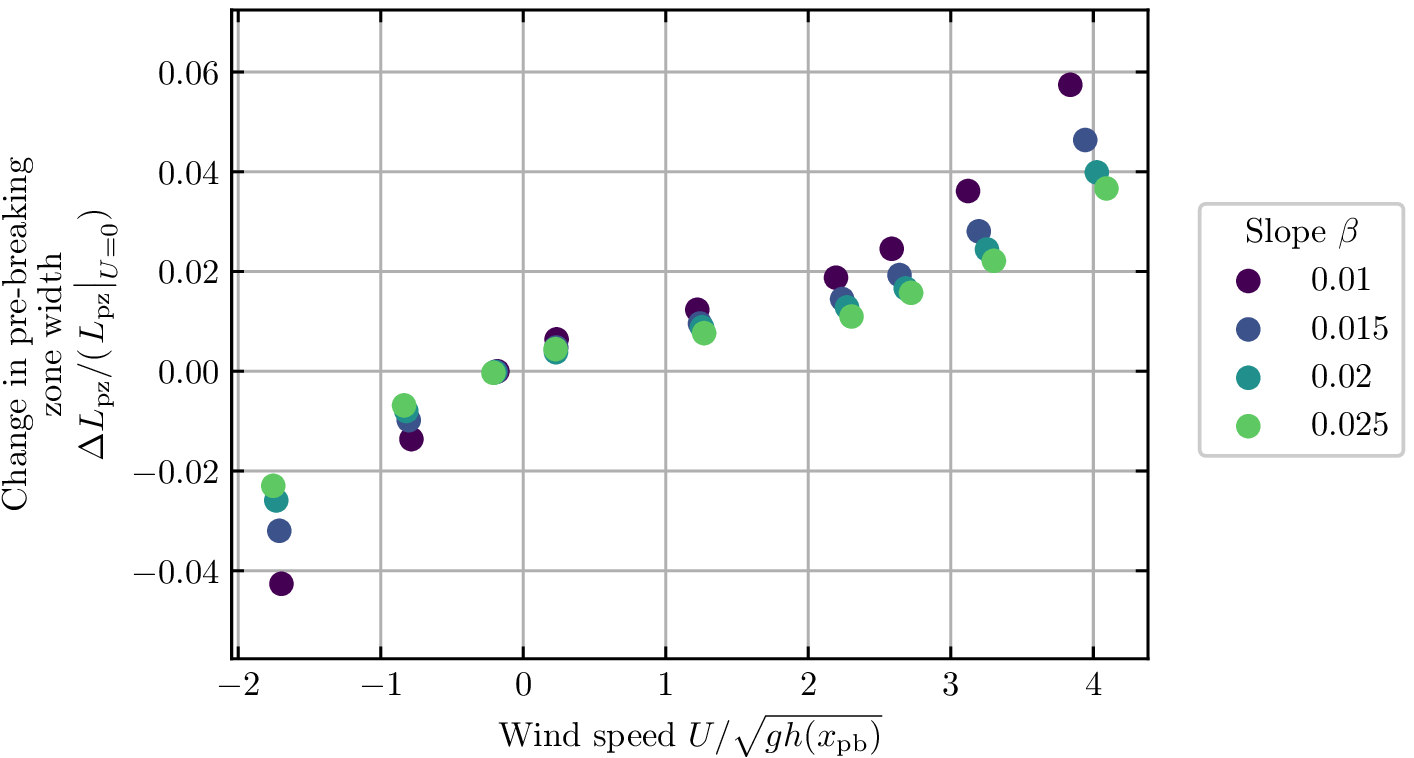}
  \caption{%
    The fractional change in prebreaking zone width $\Delta \Lpz$
    compared to the unforced case $\eval{\Lpz}_{U=0}$ (\cf{}
    \cref{sec:shape_params}) versus the nondimensional wind speed
    $U/\sqrt{g h(\xpb)}$ normalized by the local, shallow-water phase
    speed $\sqrt{g h(\xpb)}$ and evaluated at a height of half the
    solitary wave width $L$.
    The results are shown for beach slopes $\beta =
    \numrange[range-phrase={\text{--}}]{0.01}{0.025}$.
  }\label{fig:break_locations_U}
\end{figure}
We now reexamine our results regarding the prebreaking zone width
(\cref{fig:break_locations}) in terms of the wind speed $U/\sqrt{g
h(x)}$ using \cref{eq:wind_speed}.
In addition to changing the abscissa of the plot
(\cref{fig:break_locations_U}), we also modify the definition of the
change in the prebreaking zone width $\Delta \Lpz \coloneqq \Lpz -
\eval{\Lpz}_{U=0}$ by comparing and normalizing each prebreaking zone width to
the $U = 0$ case rather than the $P=0$ case.
This transformation changes the initially straight lines of
\cref{fig:break_locations} into approximate pairs of upward- and
downward-facing $\sqrt{\Delta \Lpz}$ curves shifted to the right by one
unit~(\cref{fig:break_locations_U}).
Furthermore, we see that $\Delta \Lpz$ is now much flatter for onshore
winds ($U>0$) than for equal magnitude offshore winds ($U<0$).
This is due to the inflection point of the unforced case ($P=0$) being
shifted to the right at $U/\sqrt{gh} = 1$.

\subsection{\label{sec:wind_effect} Isolating the effect of wind}
For no wind ($P=0$), solitary wave shoaling is well understood to
generate a rear shelf~\citep{miles1979korteweg}.
The variation in the rear shelf's thickness with
$P$~(\cref{fig:wave_shapes}) is reminiscent of the variability in the
wind-generated bound, dispersive, and decaying tails of flat-bottom
solitary waves~\citep{zdyrski2021wind}.
Additionally, \citet{zdyrski2021wind} showed that flat-bottom,
wind-generated tails are analogous to the dispersive tails of KdV
solutions with non-solitary-wave initial
conditions~\citep{mei2005nonlinear}.
Both the rear shelf and wind-generated tail can be viewed as weak
perturbations to the KdV equation by transforming the nondimensional
vKdV--Burgers equation \labelcref{eq:vckdv_burgers} into a
constant-coefficient, perturbed KdV equation by defining
$\nu \coloneqq (3/2) (\epsilon_0/\mu_0) \eta_0/\htil^2$
and $\tau \coloneqq \int \ctil \dd{x_1} \mu_0/(6\gamma_0)$:
\begin{equation}
  \pdv{\nu}{\tau}
  + 6 \nu \pdv{\nu}{\xi_+}
  + \pdv[3]{\nu}{\xi_+}
  =
  - \frac{9}{4} \frac{1}{\htil} \pdv{\htil}{\tau} \nu
  - 3 \frac{P_0}{\mu_0} \frac{1}{\ctil^3} \pdv[2]{\nu}{\xi_+}
  \,.
  \label{eq:pkdvb}
\end{equation}
The first term on the right-hand-side (RHS) is the shoaling term which
leads to the rear shelf~\citep{miles1979korteweg}, and the second term is
the wind-induced Burgers term~\citep{zdyrski2021wind}.
Our derivation assumed all terms in \cref{eq:pkdvb} were the same order,
and indeed $\partial_{\tau} \htil/\htil \sim 6 \gamma_0/\mu_0 =
\numrange{1.0}{2.6}$ is order one.
However, the forcing term $\abs{P_0}/\mu_0 = (4/3) \abs{P_0}/\epsilon_0
= \numrange{0}{0.07}$ is much smaller than unity and is a weak
perturbation to the sloping-bottom KdV equation with its $\sech^2$
solitary-wave and shelf solution.

Perturbed KdV equations similar to \cref{eq:pkdvb} received some
attention in past literature.
In particular, the unforced, shoaling case can be recast in a number of
asymptotically equivalent forms including
\crefor{eq:vckdv_burgers,eq:pkdvb} with $P=0$.
Previous authors applied different mathematical techniques to these
various asymptotic forms to derive closed-form approximate solutions.
For instance, \citet{knickerbocker1985propagation} solved the unforced
analog of \cref{eq:vckdv_burgers} using conservation laws.
Alternatively, \citet{grimshaw1979slowly} used a multiple-scale analysis
to solve a windless, modified form of \cref{eq:vckdv_burgers}.
Similarly, \citet{ablowitz1981solitons} solved the $P=0$ version of
\cref{eq:pkdvb} using a direct perturbation solution while
\citet{newell1985solitons} solved it using the inverse scattering
transform.
Since \cref{eq:pkdvb} shows that shoaling and wind forcing can be
treated on an equal footing, it should be possible to extend these
analyses to wind-forced solitary waves.
However, such an analysis is outside the scope of the current work, but
is suitable for future work.

\begin{figure}
  \centering
  { 
    \phantomsubfloat{\label{fig:wave_shape_differences:a}}
    \phantomsubfloat{\label{fig:wave_shape_differences:b}}
    \phantomsubfloat{\label{fig:wave_shape_differences:c}}
    \phantomsubfloat{\label{fig:wave_shape_differences:d}}
  }
  \includegraphics{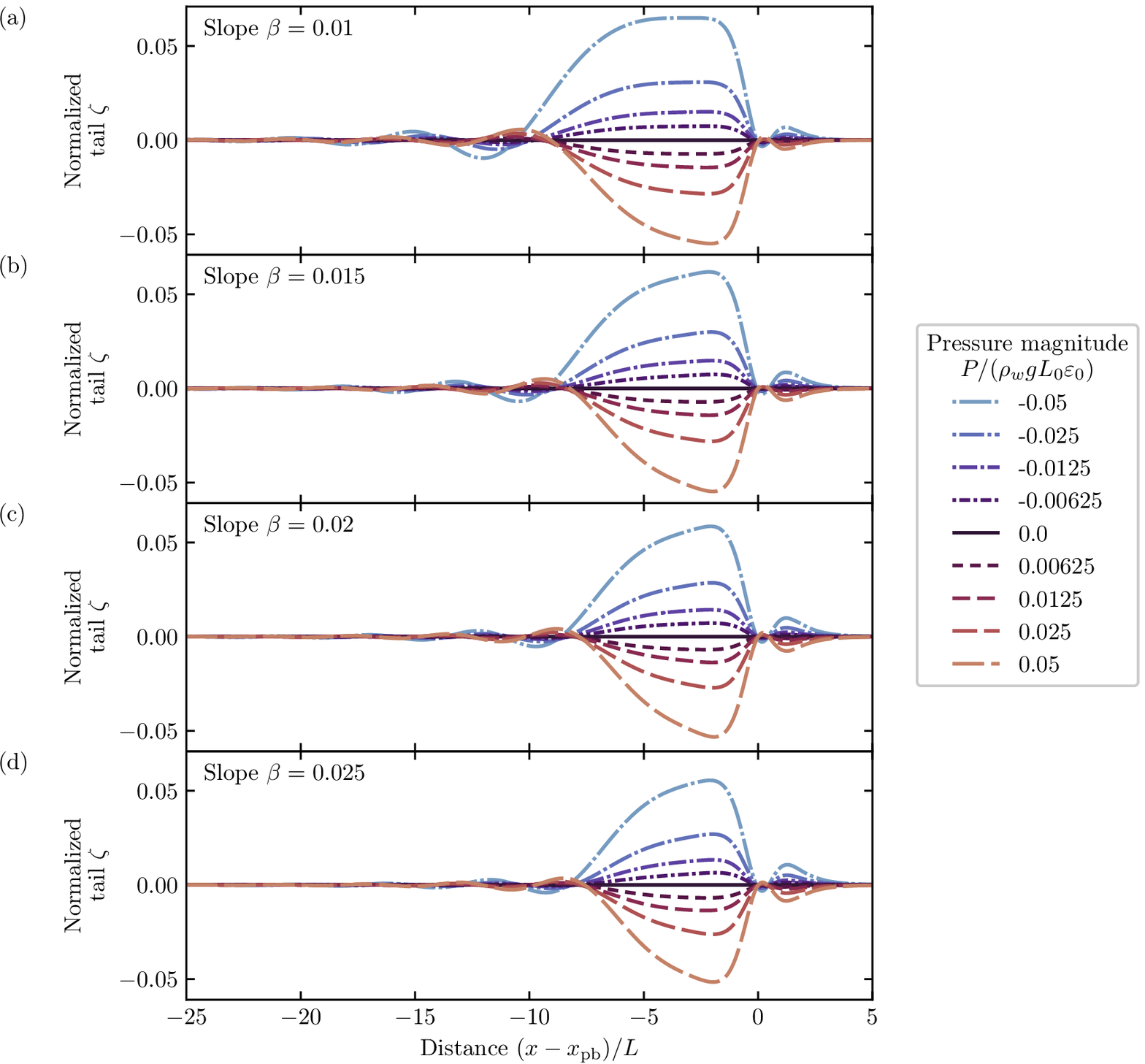}
  \caption{%
    Normalized tail $\zeta$ \cref{eq:tail_def} versus nondimensional
    position $(x-\xpb)/L$ relative to the prebreaking location $\xpb$.
    The wave profile is normalized by the maximum height $\max_x(\eta)$,
    and the spatial coordinate $(x-\xpb)$ is normalized by the wave
    width $L$ [\cref{eq:wave_width}].
    All profiles occur at
    prebreaking $\max_x(\Fr) = \Frpb = 1/3$ (\cf{}
    \cref{sec:breaking_criteria}) and are displayed for bottom slopes
    $\beta$ of
    \protect\subref{fig:wave_shape_differences:a}
    $0.01$,
    \protect\subref{fig:wave_shape_differences:b}
    $0.015$,
    \protect\subref{fig:wave_shape_differences:c}
    $0.02$,
    and
    \protect\subref{fig:wave_shape_differences:d}
    $0.025$.
    Results are shown for $\epsilon_0=0.2$, $\mu_0 = 0.15$, and
    pressure magnitude $\abs{P /(\rho_w g L_0 \epsilon_0)}$ up to
    $0.05$, as indicated in the legend.
    The solid black line is the unforced case $P = 0$ and is zero by
    definition.
  }\label{fig:wave_shape_differences}
\end{figure}

Reverting back to dimensional variables, we isolate the effect of wind
by separating out the $\sech^2$ solitary wave and the Miles rear shelf
using the unforced $P=0$ normalized profiles to represent shoaling and
rear-shelf generation.
We define a normalized tail $\zeta$ as the difference between the forced
and unforced $P=0$ normalized profiles of \cref{fig:wave_shapes}:
\begin{equation}
  \zeta \coloneqq \frac{\eta}{\max_x(\eta)} -
    \eval{\pqty{\frac{\eta}{\max_x(\eta)}}}_{P=0} \,.
  \label{eq:tail_def}
\end{equation}
For constant depth, the height $H$ and width $L$ of unforced,
$\sech^2$ solitary waves always satisfy $H L^2 = \text{const}$.
Since our numerical results (\eg \cref{fig:wave_shapes}) are dominated
by the $\sech^2$ solitary wave profile, scaling the wave profile by $H$
requires that we scale the spatial coordinate by $L \propto 1/\sqrt{H}$
to respect this symmetry and enable comparison of waves with different
heights.
We replace $h \to h(\xpeak)$ in the expression for the flat-bottom
solitary wave width $L$ \cref{eq:initial_condition_dim} to yield the
wave width for a slowly varying depth as
\begin{equation}
  L = h(\xpeak) \sqrt{\frac{4 h(\xpeak)}{3 H}} \,.
  \label{eq:wave_width}
\end{equation}
We normalize the spatial coordinate as $x/L$ to compare the normalized
tails $\zeta$ in \cref{fig:wave_shape_differences}.

We show the normalized tail $\zeta$ versus $(x-\xpb)/L$ for different
pressures $P$ and bottom slopes $\beta$ in
\cref{fig:wave_shape_differences}.
First, increasing the pressure magnitude $\abs{P}$ increases the tail's
amplitude and wavelength.
For example, the wavelength with $\beta=0.01$ is approximately $5 L$ for
$P/(\rho_w g L_0 \epsilon_0) = -0.025$ and $7.5 L$ for
$P/(\rho_w g L_0 \epsilon_0) = -0.05$.
This amplitude increase is expected, as higher pressures put more energy
into the tail, causing growth.
Additionally, increasing the bottom slope $\beta$ decreases the shelf's
width and the tail's amplitude without noticeably changing its wavelength.
We can explain the narrower shelf and smaller amplitude by recognizing
that larger $\beta$'s cause the wave to reach prebreaking (when these
profiles are compared) earlier, decreasing the time over which the wind
(tail) and shoaling (shelf) act.
The wavelength's independence of the beach slope $\beta$ also implies
that the width $L$ of the solitary wave sets the tail's wavelength.
Additionally, we note that onshore ($P>0$) and offshore ($P<0$) winds
change the polarity of the tail, consistent with
\citet{zdyrski2021wind}.
Lastly, wind induces a small, bound wave in front of the prebreaking
solitary wave with minimum near $(x-\xpb) = 0$ and extremum near
$(x-\xpb)/L \approx 2$ of the same polarity as the rear shelf
(\cref{fig:wave_shape_differences}), similar to the flat-bottom results
of \citet{zdyrski2021wind}.

Hence, the numerically calculated wave profiles (\cref{fig:wave_shapes})
are a superposition of the $\sech^2$ solitary wave, Miles'
shelf~\citep{miles1979korteweg}, and a wind-induced bound, dispersive,
and decaying tail~\citep{zdyrski2021wind}.
Furthermore, this decomposition of the full wave enables us to
understand the effects of wind and shoaling from previous studies.
The $\sech^2$ solitary wave grows and narrows due to wave
shoaling~\citep{miles1979korteweg} and
wind forcing~\citep{zdyrski2021wind}.
Miles' shelf is generated by the mass flux of the growing wave.
The shelf's absence from the normalized tails in
\cref{fig:wave_shape_differences} implies its shape is largely unchanged
by the wind, and its amplitude for a given bottom slope $\beta$ is
approximately proportional to the $\sech^2$ solitary wave.
Finally, the amplitude and wavelength of the bound, dispersive, and
decaying tail grow with the $\sech^2$ solitary
wave~\citep{zdyrski2021wind}.

This decomposition relies on the assumption that the tail and shelf are
both small compared to the solitary wave and do not influence each other
or the solitary wave.
This is only possible when the wind-forcing $P_0$ is weak and the
wave width-to-beach width ratio $\gamma_0$ is small.
\Citet{miles1979korteweg} analyzed a vKdV equation requiring the same
weak-slope assumption $\gamma_0 \sim \epsilon_0$, though his adiabatic
results required an even smaller $\gamma_0 =
\order{\num{e-2}}$~\citep{knowles2018shoaling}.
The realistic beach widths we utilized yield a $\gamma_0 =
\numrange{3e-2}{6e-2}$ somewhat larger than this adiabatic regime, and
the $\gamma_0/\epsilon_0$ term in \cref{eq:pkdvb} is not as small as the
pressure-forcing term, implying some nonlinear interactions between the
shoaling-induced shelf and the $\sech^2$ solitary are possible.
For this reason, we subtracted off the unforced solitary wave and shelf
rather than approximate them analytically.
Nevertheless, the pressure forcing $\abs{P_0}/\mu_0 =
\numrange{0}{0.07}$ we used was sufficiently small that
the weak wind forcing can be considered to interact linearly,
as seen in the clean separation between pressure-induced tail and
$\sech^2$ plus shelf in \cref{fig:wave_shape_differences}.
Thus, we show that there are physically reasonable parameter
regimes wherein wave shape and evolution are the superpositions of the
previously understood shoaling- and wind-induced effects.

\subsection{Breakpoint location comparison to prior laboratory
experiments and models}
Numerical studies investigated the effect of wind on the breaking of
shoaling solitary~\citep{xie2014numerical} and
periodic~\citep{xie2017numerical} waves using a RANS $k$--$\epsilon$
model to simulate both the air and water.
\Citet{xie2014numerical} considered solitary waves with initial height
$H_0/h_0 = 0.28$ on a beach slope of $0.05$ with onshore winds of up
to $U/\sqrt{g h_0} = 3$, while \citet{xie2017numerical} investigated
periodic waves with initial height $H_0/h_0 = 0.3$ and initial
inverse wavelength $h_0/\lambda_0 = 0.1$ on a beach slope of $0.03$
forced by onshore winds up to $U/\sqrt{g h_0} = 2$.
These studies determined that the (absolute) maximum wave heights
$\max_t(\eta)/h_0$ increased with increasing onshore wind at each
location $x<\xb$ prior to breaking at $\xb$, consistent with our
findings in \cref{fig:statistics_solitary:b}.
Furthermore, \citet{xie2014numerical} found the effect of wind on
breaker depth is significant while the effect on breaker height is
minor, again consistent with our prebreaking findings.
Finally, comparing wave profiles in \citet{xie2014numerical} shows that
onshore winds increase the wave slope at a fixed location, which is
consistent with our \cref{fig:statistics_solitary:c}.

For periodic waves, previous laboratory experiments also investigated
wind's effect on the breaking characteristics of shoaling
waves~\citep{douglass1990influence,king1996changes}.
\Citet{douglass1990influence} considered waves with initial height
$H_0/h_0 = 0.3$ and initial inverse wavelength $h_0/\lambda_0 = 0.1$
under wind speeds of up to $U/\sqrt{g h_0} = \pm 2.3$ on a beach with
slope $0.04$ while \citet{king1996changes} considered waves with initial
height $H_0/h_0 = 0.2$ and initial inverse wavelength $h_0/\lambda_0 =
0.3$ with wind speeds of up to $U/\sqrt{g h_0} = \pm 1.1$ on a beach
with slope $0.05$.
\Citet{douglass1990influence} measured how wind speed affects wave
parameters and changes the surf zone width for periodic waves.
Directly comparing our \cref{fig:break_locations_U} to \figname{} 2 of
\citet{douglass1990influence}, we see many qualitative
similarities, including the prebreaking zone width's flatter response
near $U=0$ and a stronger response for offshore winds ($U<0$) than the
corresponding onshore winds ($U>0$), with our change roughly four times
smaller than theirs.
Furthermore, the laboratory studies also found that the relative
breaking height $\Hb/\hb$, normalized by the breaking depth,
decreased by as much as \SI{40}{\percent} for offshore wind speeds of
$U/\sqrt{g\hb} = 4$ and increased by up to \SI{10}{\percent} for
onshore wind speeds of $U/\sqrt{g\hb} = -2$ compared to the unforced
case~\citep{douglass1990influence,king1996changes}.
By comparison, over those same wind speed ranges of $U/\sqrt{g \hpb}
= 1 \pm 3$, our simulations found that the relative prebreaking height
$\Hpb/\hpb$ varied by approximately \SI{1}{\percent} between
onshore and offshore winds [\cref{fig:statistics_solitary:b}], with the
same polarity as the laboratory experiments.
Finally, \citet{douglass1990influence} observed only a slight
dependence of the breaking wave height on wind speed, which is
consistent with our finding that offshore-forced waves are only
\SI{1}{\percent} larger than onshore-forced waves at prebreaking.
On the contrary, \citet{king1996changes} measured no statistically
significant change with wind speed in the ratio between the change in
the fractional change $(\Hb - \eval{\Hb}_{U=0})/\eval{\Hb}_{U=0}$ of
(absolute) breaking height $H_b$ compared to the unforced case
$\eval{\Hb}_{U=0}$.
This is surprising, as both our results and those of
\citet{douglass1990influence} had near-constant relative (pre)breaking
height $\Delta(\Hb/\hb)$.
Indeed, constant $\Delta(\Hb/\hb)$ implies
$\Delta \Hb/\eval{\Hb}_{U=0}$ must vary with wind (we find approximately
$\pm \SI{5}{\percent}$ at prebreaking) to counteract the varying of
$\Delta \hb/\eval{\hb}_{U=0} = \Delta L_b/\eval{L_b}_{U=0}$ with wind
that \citet{douglass1990influence}, \citet{king1996changes}, and we all
find.

Our results qualitatively agree with prior numerical results on solitary
waves~\citep{xie2014numerical} as well as experimental and numerical
results on periodic
waves~\citep{douglass1990influence,king1996changes,xie2017numerical},
and the quantitative mismatch can be partly explained by the different
nondimensional parameters.
All three studies mentioned used larger initial waves ($\epsilon_0
\approx 0.3$), so nonlinear effects were likely more important.
They also used steeper beach slopes, enhancing the shoaling effect.
Additionally, while the surf zone width change
is roughly four times larger for \citet{douglass1990influence} than for
our simulations over the same wind-speed range,
\citet{douglass1990influence} investigated waves that were actually
breaking.
In contrast, we stopped our simulations at prebreaking $\max_x(\Fr) =
\Frpb = 1/3$, significantly before actual breaking $\max_x(\Fr) \approx
1$~\citep{derakhti2020unified}, thus we expect smaller changes to the
fractional surf prebreaking zone width (\cf{}
\cref{sec:pb_width_results}).

\subsection{Wave shape comparison to prior laboratory experiments}
\begin{figure}
  \centering
  { 
    \phantomsubfloat{\label{fig:fv2005_comparison:a}}
    \phantomsubfloat{\label{fig:fv2005_comparison:b}}
  }
  \includegraphics{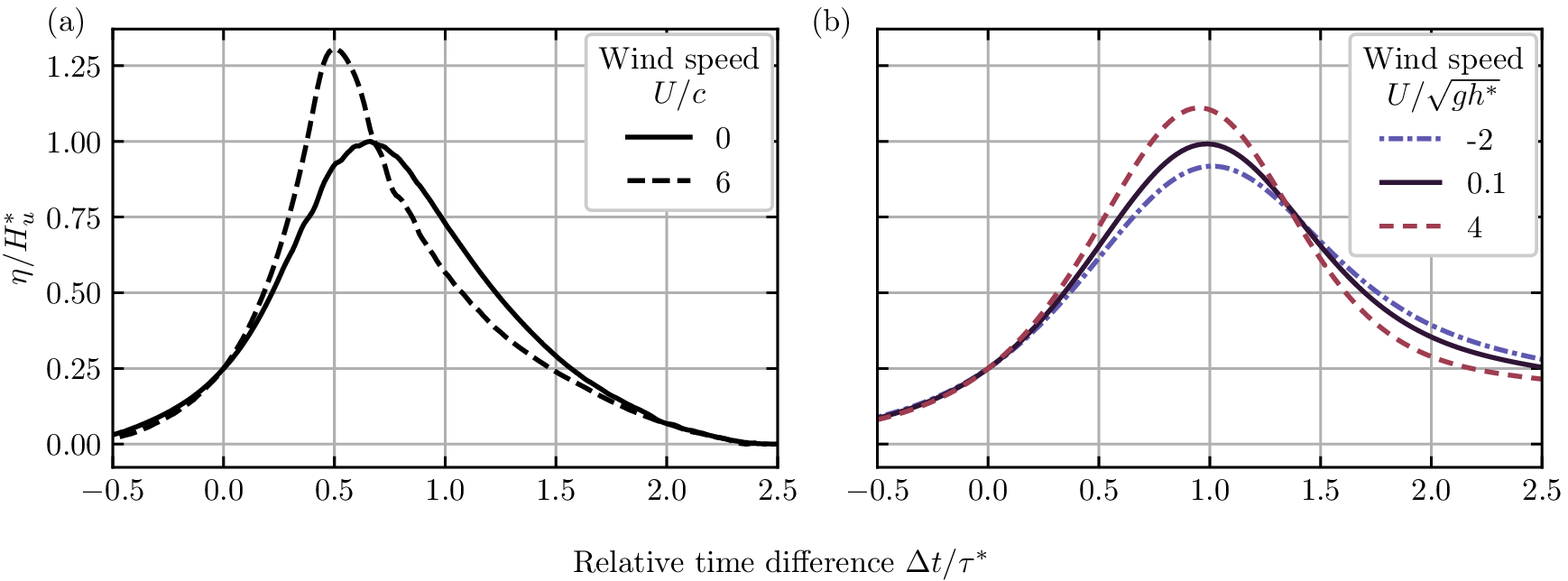}
  \caption{%
  Nondimensional wave profile $\eta/\HnoWind$ at a fixed location
  versus nondimensional time $\Delta t/\tau^*$ with $\eta$ normalized
  by the no-wind wave height $\HnoWind$.
  The time difference $\Delta t \coloneqq t-\tquarter$ is relative to the
  time when the profile reaches $\eta/\HnoWind = 0.25$ and is normalized
  by the unforced solitary wave temporal width $\tau^* =
  \LnoWind/\sqrt{gh^*}$ at a depth $h^*$, with $\LnoWind$ the unforced
  spatial width [\cref{eq:wave_width}].
  \protect\subref{fig:fv2005_comparison:a}
  Periodic wave laboratory results from \citet{feddersen2005wind} with
  $\epsilon_0 \approx 0.28$, $\mu_0 = 0.035$, and $\beta = 0.125$.
  The solid line represents the no-wind case $U/c=0$ and the dashed
  line corresponds to an onshore wind $U/c=6$, as indicated in the
  legend.
  The waves began shoaling at a depth of $h_0 = \SI{0.37}{\meter}$ and
  were measured at a nondimensional depth $h^*/h_0 = 0.62$.
  The wave height $\eta$ is plotted relative to the wave trough
  $\min_t(\eta)$.
  \protect\subref{fig:fv2005_comparison:b}
  Numerical results for $\epsilon_0 = 0.2$, $\mu_0 = 0.15$, and $\beta =
  0.015$ at a nondimensional water depth $h^*/h_0 = 0.69$ and for
  nondimensional wind speeds $U/\sqrt{gh^*}$ [\cf{}
  \cref{eq:wind_speed}] as indicated in the legend.
  }\label{fig:fv2005_comparison}
\end{figure}

\Citet{feddersen2005wind} experimentally examined the effect of wind on
the temporal shape of shoaling, periodic waves at a fixed location for
no-wind ($U/c=0$) and onshore-wind ($U/c=6$) cases [\cf{}
\cref{fig:fv2005_comparison:a}].
The waves began shoaling at a depth of $h_0=\SI{0.37}{\meter}$ and were
observed at a depth $h^*/h_0 = 0.62$.
To enable direct comparison with our results, we normalized the profile
by the height of the no-wind profile $\HnoWind$ at depth $h^*$.
We also normalized the time coordinate with an unforced solitary wave's
temporal width $\tau^* = \LnoWind/\sqrt{gh^*}$ at depth $h^*$, where
the unforced spatial width $\LnoWind$ is related to the wave height via
an equation similar to \cref{eq:wave_width}.
We observe that the onshore wind increases the wave height and causes
earlier peak arrival, relative to the arrival time of $\eta/\HnoWind =
0.25$ (\cref{fig:fv2005_comparison}).
For comparison, our numerical results [\cref{fig:fv2005_comparison:b}]
over different wind speeds, calculated using \cref{eq:wind_speed}, are
shown at $x/h = 20.1$ corresponding to $h^*/h_0=0.68$.
This corresponds to $\xpb$ for the strongest onshore wind $U/\sqrt{gh^*} =
4$ (\ie $P=0.05$) case.
We truncated the time series at $\Fr = 1/2$, instead of $\Frpb = 1/3$,
to enable qualitative comparison of the strong-wind numerical case
($U/\sqrt{gh^*}=4$) with the laboratory onshore-wind case ($U/c=6$).
However, we note that this result is strictly out of the asymptotic
range.

Our numerical results [\cref{fig:fv2005_comparison:b}] are qualitatively
similar to the experimental results of \citet{feddersen2005wind}
[\cref{fig:fv2005_comparison:a}], despite different experimental
and model conditions.
Specifically, \citet{feddersen2005wind} presented shallower
($h^*/h_0=0.62$) measurements of shoaling periodic waves ($h_0/\lambda_0
= 0.27$) forced by stronger laboratory winds ($U/c = 6$) in contrast to
our deeper ($h^*/h_0=0.68$) simulations of solitary waves
($h_0/\lambda_0 \to 0$) for weaker winds ($U/\sqrt{gh^*} = 4$).
Therefore, we display the results side by side as opposed to overlaid to
emphasize the qualitative comparison.
Onshore wind causes wave growth and narrowing for both laboratory
observations and numerical results.
The peaks of the onshore-forced waves occur earlier (relative to the
time of $\eta/\HnoWind=0.25$) due to wave narrowing, as seen in
\cref{fig:statistics_solitary:d}.
Finally, the rear faces of the onshore-forced waves
($\Delta t/\tau^*$ from \numrange[range-phrase = \text{~to~}]{1}{2.5})
dip below the no-wind cases.
Numerical results with offshore wind continue the pattern.
The qualitative similarities between our results and that of
\citet{feddersen2005wind} suggests that the our vKdV--Burgers equation
captures the essential aspects of wind-induced effects on shoaling wave
shape.
No other theoretical model has yet shown such qualitative similarity
with wind-forced wave shape experiments, despite the differences
(\eg periodic versus solitary) between the laboratory and numerical
studies.

\section{Conclusion}
\noindent
While shoaling-induced changes to wave shape are well understood,
the interaction of wind-induced and shoaling-induced shape changes has
been less studied and lacked a theoretical framework.
Utilizing a Jeffreys-type wind-induced surface pressure, we defined four
nondimensional parameters that controlled our system: the initial wave
height $\epsilon_0$, the inverse wavelength squared $\mu_0$, the
pressure strength $P_0$, and the wave width-to-beach width ratio $\gamma_0$.
We leveraged these small parameters to reduce the forced,
variable-bathymetry Boussinesq equations to a variable-coefficient
Korteweg--de Vries--Burgers equation for the wave profile $\eta$.
We also extended the convective breaking criterion of
\citet{brun2018convective} to include pressure and shoaling.
A third-order Runge-Kutta solver determined the time evolution of a
solitary wave initial condition up a planar beach under the influence of
onshore and offshore winds.
Stopping the simulations at a prebreaking Froude number of $1/3$
revealed that the prebreaking relative height and maximum slope are
largely independent of wind speed, but onshore winds cause a narrowing
of the waves.
The width of the prebreaking zone is strongly modulated by wind speed,
with offshore wind decreasing the prebreaking zone width by
approximately \SI{5}{\percent} for the mildest beach slopes.
Investigating the wave shape at prebreaking revealed that the front of
the wave is relatively unchanged and matches an unforced solitary wave,
while the rear shelf is strongly affected by wind speed and bottom
slope.
We isolated the effect of wind from the effect of shoaling and revealed
a bound, dispersive, and decaying tail similar to wind-induced tails on
flat bottoms.
By leveraging the relationship between surface pressure $P$ and wind
speed $U$, we directly compared our results to existing experimental and
numerical results.
We found qualitative agreement in surf width changes and wave height
changes and expect better quantitative agreement as the waves propagate
closer to breaking.
These results suggest that wind significantly impacts wave breaking, and
our simplified model highlights the relevant physics and changes to wave
shape.
Future avenues of research could include calculating asymptotic, closed
form solutions to \cref{eq:pkdvb} or deriving coupled equations for both
the water and air motions to more accurately predict the surface
pressure distribution.

\begin{acknowledgments}
\noindent
We are grateful to D.~G.~Grimes and M.~S.~Spydell for discussions on
this work.
Additionally, we thank D.~P.~Arovas, J.~A.~McGreevy, P.~H.~Diamond, and
W.~R.~Young for their invaluable insights and recommendations.
Furthermore, we thank J.~Knowles for assisting our comparisons to
his results.
We thank the National Science Foundation (OCE-1558695) and the Mark Walk
Wolfinger Surfzone Processes Research Fund for their support of this
work.
\end{acknowledgments}

\end{document}